\documentclass[useAMS,usenatbib,fleqn]{mnras}
\usepackage{multirow}
\usepackage{graphicx}
\usepackage{color,soul}
\usepackage{graphicx}
\usepackage{subfig}
\usepackage{epsfig}
\usepackage{epstopdf}
\usepackage{amssymb,amsmath}
\usepackage{aas_macros}
\usepackage[T1]{fontenc}

\title[Spectral Indices]{
Probing Plasma Physics with Spectral Index Maps of Accreting Black Holes on Event Horizon Scales
}

\author[Ricarte et al.]{Angelo Ricarte$^{1,2}$, Charles Gammie$^{3,4}$, Ramesh Narayan$^{1,2}$, and Ben S.~Prather$^5$
\\
$^{1}$ Center for Astrophysics | Harvard \& Smithsonian, 60 Garden Street, Cambridge, MA 02138, USA \\
$^{2}$ Black Hole Initiative at Harvard University, 20 Garden Street, Cambridge, MA 02138, USA \\
$^{3}$ Department of Physics, University of Illinois at Urbana–Champaign, 1110 West Green Street, Urbana, IL 61801, USA \\
$^{4}$ Department of Astronomy, University of Illinois at Urbana–Champaign, 1002 West Green Street, Urbana, IL 61801, USA \\ 
$^{5}$ CCS-2, Los Alamos National Laboratory, P.O. Box 1663, Los Alamos, NM 87545, USA
}

\date{\today}

\begin{document}
\pagerange{\pageref{firstpage}--\pageref{lastpage}} \pubyear{2022}
\maketitle

\begin{abstract}
The Event Horizon Telescope (EHT) collaboration has produced the first resolved images of the supermassive black holes at the centre of our galaxy and at the centre of the elliptical galaxy M87.  As both technology and analysis pipelines improve, it will soon become possible to produce spectral index maps of black hole accretion flows on event horizon scales.  In this work, we predict spectral index maps of both M87* and Sgr A* by applying the general relativistic radiative transfer (GRRT) code {\sc ipole} to a suite of general relativistic magnetohydrodynamic (GRMHD) simulations.  We analytically show that the spectral index increases with increasing magnetic field strength, electron temperature, and optical depth.  Consequently, spectral index maps grow more negative with increasing radius in almost all models, since all of these quantities tend to be maximised near the event horizon.  Additionally, photon ring geodesics exhibit more positive spectral indices, since they sample the innermost regions of the accretion flow with the most extreme plasma conditions.  Spectral index maps are sensitive to highly uncertain plasma heating prescriptions (the electron temperature and distribution function).  However, if our understanding of these aspects of plasma physics can be tightened, even the spatially unresolved spectral index around 230 GHz can be used to discriminate between models.  In particular, Standard and Normal Evolution (SANE) flows tend to exhibit more negative spectral indices than Magnetically Arrested Disk (MAD) flows due to differences in the characteristic magnetic field strength and temperature of emitting plasma.
\end{abstract}


\section{Introduction}
\label{sec:introduction}

Every massive galaxy is believed to host a supermassive black hole (supermassive BH; SMBH) at its centre that, when supplied with gas to accrete, can shine and impart energy into its host galaxy as an active galactic nucleus (AGN) \citep{Kormendy&Richstone1995,Kormendy&Ho2013}.  ``AGN feedback'' in the form of radiation, winds, and/or jets is believed to be essential for regulating gas cooling in massive galaxies and is therefore a critical piece of galaxy evolution modelling \citep[e.g.,][]{Croton+2006,Heckman&Best2014}.  Yet at present, the details of both accretion and feedback processes are poorly understood, motivating in-depth studies of the central engines of AGN.

In the past few years, the Event Horizon Telescope (EHT) has produced the first resolved images of SMBHs, ushering in a new era of spatially resolved accretion physics on event horizon scales.  These include the SMBH at the centre of the M87 galaxy, henceforth M87* \citep{EHT1,EHT2,EHT3,EHT4,EHT5,EHT6,EHT7,EHT8}, as well as the SMBH at the centre of our own galaxy, Sagittarius A* or Sgr A* \citep{EHT_Sgra_I,EHT_Sgra_II,EHT_SgrA_III,EHT_SgrA_IV,EHT_SgrA_V,EHT_SgrA_VI}.  Each is believed to be enveloped in a radiatively inefficient accretion flow (RIAF), a hot, geometrically thick, and optically thin accretion disk with a low Eddington ratio \citep{Narayan&Yi1994,Narayan&Yi1995,Abramowicz+1995,Narayan+2000,Quataert&Gruzinov2000a,Yuan&Narayan2014}.  For these systems, two of the most interesting parameters to constrain observationally are its spin and magnetic field state.  The spin of a SMBH can be tapped to power jets \citep{Blandford&Znajek1977}, and its cosmic evolution is sensitive to uncertain SMBH accretion and dynamics over billions of years \citep[e.g.,][]{King+2008,Volonteri+2013}.  The magnetic field state of the accretion flow, whether a ``Magnetically Arrested Disk'' (MAD) or ``Standard and Normal Evolution'' (SANE) (as discussed further in \autoref{sec:grmhd}) describes whether or not the magnetic field is ordered and strong enough to affect the plasma dynamics.  EHT imaging with resolved polarimetry, combined with other multi-wavelength constraints, currently favour a MAD model with non-zero spin for M87* \citep{EHT5,EHT8}.  For Sgr A*, a promising cluster of relatively face-on, spinning MAD models passes most observational constraints \citep{EHT_SgrA_V}, and a full polarimetric analysis is underway.

The EHT aims to continue monitoring these sources while improving its capabilities by adding sites, improving bandwidth and recording speed, and enabling observations at other frequencies such as 86 and 345 GHz \citep{Doeleman+2019,Raymond+2021}.  Continued monitoring alone is likely to constrain models based on time variability, and structural variability on event horizon scales has already been detected in EHT and ``proto-EHT'' data \citep{Wielgus+2020}.  In addition to simply increasing the signal to noise, expanding the observed bandwidth can enable both total intensity and polarimetry studies in the frequency domain.  Resolved rotation measure maps may reveal turbulent and non-uniform magnetic field structures that are predicted in simulations \citep{Ricarte+2020}.  It will also be possible to produce resolved spectral index maps on event horizon scales, which is the subject of this work.  The spectral index of an emitting plasma, which we will denote as $\alpha \equiv d\log I/d\log \nu$, where $I$ is the intensity and $\nu$ is the wavelength, is sensitive to its underlying magnetic field, its optical depth, and its electron energy distribution \citep[e.g.,][]{Rybicki&Lightman1986}, all of which vary substantially in our models.

We focus our study on 230 GHz, the primary observing frequency of the EHT.  As observed by the Atacama Large Millimeter Array (ALMA), Sgr A* exhibits $\alpha \approx 0.0 \pm 0.1$ at approximately this frequency \citep[]{Goddi+2021,Wielgus+2022}, with a more negative spectral index of $\alpha \approx -0.3$ between 230 GHz and 870 GHz \citep{Bower+2019}.  For the core of M87, VLBI observations determined $\alpha \approx -0.4$ between 15 GHz and 129 GHz, although the true spectral index may be nearer to 0 and steepened due to changes in (u,v) coverage \citep{Kim+2018}.  Meanwhile, $\alpha \approx -1.1$ was observed at 230 GHz using the Atacama Large Millimeter Array (ALMA) alone, but this measurement includes larger scale jet emission which probably drives the spectral index downwards \citep{Goddi+2021}.  Interpreting these significantly different spectral indices for M87 is complicated by the existence of extended emission that is resolved out by the EHT, as well as differences in (u,v) coverage, as discussed in Appendix A of \citet{Kim+2018}.

In this work, we perform an exhaustive study of the spectral index on event horizon scales both analytically and using ray-traced general relativistic magnetohydrodynamics (GRMHD) simulations.  In \autoref{sec:methodology}, we describe the GRMHD and ray-tracing calculations upon which most of this work is based.  In \autoref{sec:analytic}, we briefly explore analytic expectations of the spectral index for a uniform slab of plasma.  In \autoref{sec:results}, we discuss the main results of our simulated images, which include both spatially resolved spectral index maps and spatially unresolved spectral indices.  We discuss further implications of our results in \autoref{sec:discussion}, then summarise our conclusions in \autoref{sec:conclusion}.

\section{Methodology}
\label{sec:methodology}

\subsection{GRMHD}
\label{sec:grmhd}

As our starting point, we consider ten general relativistic magnetohydrodynamics (GRMHD) simulations, first presented in \citet{EHT_SgrA_V}.  The simulations considered were run with the GRMHD code {\sc KHARMA} \citep[Prather et al. in prep.,][]{Prather+2021}.  The ten GRMHD models cover two types of accretion states, MAD and SANE, around BHs with five different dimensionless spin parameters $a_\bullet \in \{-0.94, -0.5, 0, 0.5, 0.94 \}$.  MAD models accumulate strong and ordered magnetic fields that are powerful enough to affect disk dynamics \citep{Bisnovatyi-Kogan&Ruzmaikin1974,Igumenshchev+2003,Narayan+2003,Chael+2019}, while SANE models have dynamically weaker and more turbulent magnetic fields \citep{Narayan+2012,Sadowski+2013,Ryan+2018}.

All models are performed at a resolution of 288x128x128 zones in $r$, $\theta$, and $\phi$ directions respectively, on a domain of radius 1000 $GM_\bullet/c^2$.  The simulations in this work use the formulae and “funky” MKS coordinate system and grid sizing outlined in \citet{Wong+2022}.  Simulations are performed using RK2 time-stepping and WENO5 spatial reconstruction, with code details as outlined in \citet{Prather+2021}.  Plasma is initialised in a \citet{Fishbone&Moncrief1976} equilibrum torus solution, magnetised proportionally to density in SANEs and to both the density and the cube of radius in MADs.  MADs are initialised with a torus with an inner boundary at $20 \; GM_\bullet/c^2$ and a pressure maximum at $40 \; GM_\bullet/c^2$, while SANEs are initialised with a torus with an inner boundary at $10 \; GM_\bullet/c^2$ and a pressure maximum at $20 \; GM_\bullet/c^2$.  Magnetic flux is initialised and normalised as described in \citet{Wong+2022} equations 33 and 34.  All models are run to a final time of $3 \times 10^4 \ GM_\bullet/c^3$, where $G$ is the gravitational constant, $M_\bullet$ is the black hole mass, and $c$ is the speed of light.  To reduce the number of images needed for this study but still obtain some sense of the time variability, we image only eleven snapshots from each model, spaced evenly between $2.5\times 10^4  \ GM_\bullet/c^3$ and $3.0\times 10^4  \ GM_\bullet/c^3$.  Since ideal GRMHD simulations are scale-free, we are able to use the same simulations to produce images of both M87* and Sgr A*.

\subsection{GRRT}
\label{sec:grrt}

We use the code {\sc ipole} to perform General Relativistic Ray Tracing (GRRT) calculations on the GRMHD outputs \citep{Moscibrodzka&Gammie2018}.  At this step, we specify the BH mass, its distance, the mass units of the plasma, and parameters related to electron heating.  For M87*, we assume a BH mass of $6.2 \times 10^9 \ \mathrm{M}_\odot$ and a distance of 16.9 Mpc from stellar dynamics \citep{Gebhardt+2011}, for consistency with \citet{EHT5,EHT8}.  For Sgr A*, we assume a BH mass of $4.14 \times 10^6 \ \mathrm{M}_\odot$ and a distance of 8.123 kpc from stellar orbit studies \citep{Boehle+2016}.  GRMHD simulations are invariant under the transformation $\rho \to \mathcal{M} \rho$ and $B \to \sqrt{\mathcal{M}} B$, where $\rho$ is the mass density, $B$ is the magnetic field strength, and $\mathcal{M}$ is a scalar.  Thus, we fit $\mathcal{M}$ such that at 228 GHz models of M87* maintain an average flux of 0.5 Jy and models of Sgr A* maintain an average flux of 2.4 Jy \citep{Goddi+2021}.  For a given model, this fixes the accretion rate, which is not a free parameter in our work.

The diffuse plasma accreting onto Sgr A* and M87* have low enough densities such that the mean free path is much larger than the size of the system.  As a result, ions and electrons are not believed to achieve thermal equilibrium with each other \citep{Shapiro+1976,Rees+1982,Narayan+1995}.  Models of electron heating predict that electrons are heated less efficiently than ions, and therefore acquire lower temperatures.  The temperature ratio depends on the plasma $\beta \equiv P_\mathrm{gas}/P_\mathrm{mag}$, the ratio of the gas to magnetic pressure.  More equal rates of heating are obtained when $\beta \to 0$, while the ions acquire more of the heating when $\beta > 1$  \citep{Howes2010,Kawazura+2019,Chael+2019,Rowan+2019,Mizuno+2021}.  To account for this effect, we prescribe the electron temperature relative to the ion temperature (which originates from the GRMHD) via the widely-used \citet{Moscibrodzka+2016} prescription,

\begin{equation}
    \frac{T_i}{T_e} = R_\mathrm{high} \frac{\beta^2}{1+\beta^2} + R_\mathrm{low} \frac{1}{1+\beta^2}.
\end{equation}

\noindent  For this study, we consider $R_\mathrm{high} \in \{1,10,40,160\}$ but fix $R_\mathrm{low}=1$.

While most GRRT in the literature is performed assuming a thermal electron distribution function (eDF), the spectral energy distributions of Sgr A* and M87* favour a non-thermal component.  Higher energy electron populations than available in a thermal eDF are believed to be important for reproducing the flux in the near-infrared and can increase the sizes of images \citep{Ozel+2000,Mao+2017,Davelaar+2018}.

For our non-thermal eDF models, we explore three different ``kappa'' distributions, which have a thermal core and a power law tail that extends to high energies with slope $d\log n/d\log \gamma = \kappa-1$ \citep{Vasyliunas1968}.  This distribution finds its origins in fits to the observed solar wind \citep{Decker&Krimigis2003,Pierrard&Lazar2010}, and is naturally produced in simulations of particle acceleration in an accretion disk \citep{Kunz+2016}.  The relativistic kappa distribution is given by

\begin{equation}
    \frac{dn_e}{d\gamma d\cos \xi d\phi} = \frac{N}{4\pi}\gamma(\gamma^2-1)^{1/2}\left(1+\frac{\gamma-1}{\kappa w}\right)^{-(\kappa+1)},
\end{equation}

\noindent where $N$ is the overall normalisation, $\gamma$ is the Lorentz factor, $\xi$ is the pitch angle, $\phi$ is the gyrophase, and $\kappa$ and $w$ are dimensionless free parameters \citep{Xiao2006}.  As $\kappa \to \infty$, the kappa distribution becomes a thermal distribution.  For our study, we explore three global values for $\kappa \in \{ 3.5, 5.0, 7.0\}$ that do not change within the simulated region.  Meanwhile, we scale the width $w$ with the electron temperature in each cell via

\begin{equation}
    w = \Theta_e \frac{\kappa-3}{\kappa},
\end{equation}

\noindent where $\Theta_e$ is the electron temperature in units of the electron rest mass.  This prescription preserves the total amount of energy in the distribution with respect to thermal (as long as $\kappa<3$).

We adopt the polarized emissivities $j_\nu$ and absorptivities $\alpha_\nu$ for a kappa distribution from {\sc Symphony} \citep{Pandya+2016}.  {\sc Symphony} provides fitting formulae for $j_\nu$ and $\alpha_\nu$ as a function of magnetic field, electron temperature, and electron number density.  Meanwhile, our radiative transfer coefficients for thermal distributions originate from \citet{Dexter2016}.  We set all radiative transfer coefficients to 0 in highly magnetized regions ($\sigma \equiv B^2/\rho > \sigma_\mathrm{cut} = 1$), where numerical floors can result in the artificial injection of material.  We do not expect that this cut has strong qualitative effects on our results, although it may be more impactful in models where $\kappa$ scales with $\sigma$ \citep[e.g.,][]{Davelaar+2018,Fromm+2022,Scepi+2022}.  The effect of this choice is briefly explored in Appendix \ref{sec:sigma_cut}.

In summary, we create images for 2 sources (M87*, Sgr A*), 2 magnetic field states (MAD, SANE), 5 spins, 11 snapshots, 4 values of $R_\mathrm{high}$, 4 eDF models, and for Sgr A* 3 inclinations.  For each of these models, we produce images at 4 frequencies, $\nu \in \{214.1, 228.1 \}$ GHz for Sgr A* and $\nu \in \{214.1, 228.1, 340.0, 350.0 \}$ GHz for M87*, where the first two values are specifically chosen to emulate EHT 2018 observations.  This results in a library containing a total of 17,600 images.  These are created with an angular resolution of 0.5 $\mu$as, and a field of view of 240 $\mu$as for M87* and 300 $\mu$as for Sgr A*.  These are larger fields of view than typically adopted, since images with non-thermal electrons tend to be larger than thermal ones \citep{Ozel+2000,Mao+2017,Davelaar+2018}.  For all images, an outer integration radius of $50 \ G M_\bullet/c^2$ is used for the radiative transfer.

Throughout, we define the spectral index $\alpha \equiv d\log I_\nu/d\log \nu$.  When considering an image's spectral index, $\alpha_\mathrm{net}$ is the spectral index one would infer from a spatially unresolved measurement.

\section{Analytic Expectations}
\label{sec:analytic}

To interpret the results of our GRRT calculations, let us first explore analytic calculations for a uniform slab of material with a magnetic field strength $|B|$, electron temperature in units of rest mass energy $\Theta_e$, and thickness $s$.  Throughout, we also assume that the magnetic field is oriented perpendicular to the photon wave-vector.  From these properties, we can compute the emissivity $j_\nu(B,\Theta_e)$ and absorptivity $\alpha_\nu(B,\Theta_e)$\footnote{Note that we use $\alpha_\nu$ to represent the absorptivity, but $\alpha$ to represent the spectral index.} using the fitting functions provided in \citet{Pandya+2016}.  With this simple setup, the radiative transfer equation reduces to

\begin{equation}
    I_\nu = \frac{j_\nu}{\alpha_\nu}\left( 1-e^{-\tau_\nu} \right), \
    \tau_\nu \equiv \alpha_\nu s, 
    \label{eqn:radiative_transfer_slab}
\end{equation}

\noindent where $I_\nu$ is the observed intensity, and $\tau_\nu$ is the optical depth.  

\subsubsection{Optically Thin}

\begin{figure}
  \centering
  \includegraphics[width=0.5\textwidth]{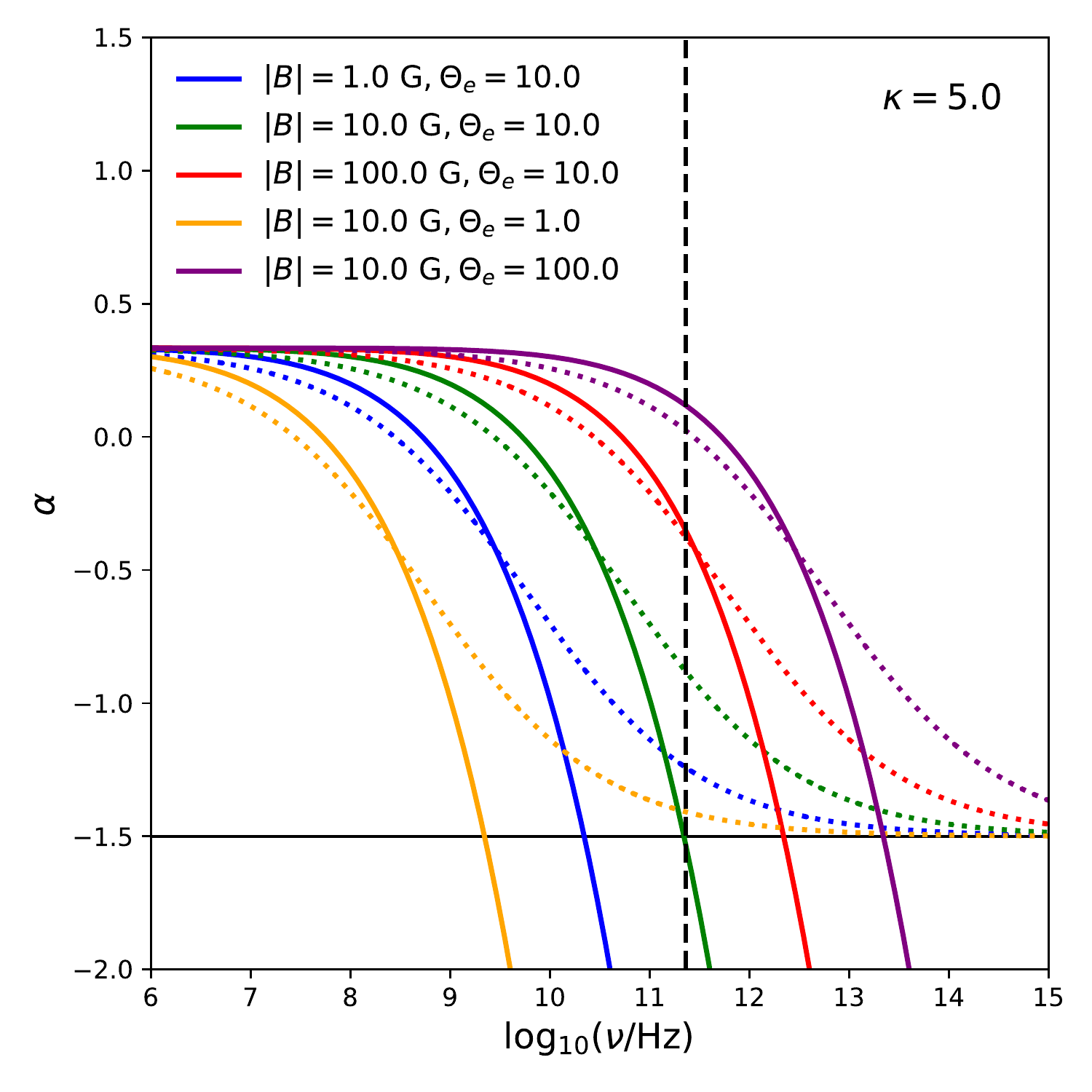}
  \caption{Spectral index as a function of frequency in the optically thin limit for a uniform slab of plasma.  Models with $\kappa=5$ are plotted with dotted lines, while thermal models are plotted with solid lines.  Different colours correspond to different combinations of magnetic field strength $|B|$ and electron temperature $\Theta_e$.  A black vertical dashed line marks 230 GHz, the primary observing frequency of the EHT, and the focus of this study.  In the optically thin and low frequency limit, both classes of models approach $\alpha=1/3$.  Models turn over at a critical frequency $\nu_\mathrm{crit} \propto B\Theta_e^2$.  In the optically thin and high frequency limit, kappa models asymptote to a value of $\alpha_\mathrm{min}=-(\kappa-2)/2$, plotted with a horizontal black line, while thermal models diverge to $-\infty$.  \label{fig:analytic_optically_thin}}
\end{figure}

In the optically thin limit ($\tau_\nu \to 0$), \autoref{eqn:radiative_transfer_slab} reduces to $I_\nu = j_\nu s$. Thus, the spectral index depends only on the emissivity coefficient $j_\nu$, via $\alpha = d\log j_\nu / d\log \nu$.  In \autoref{fig:analytic_optically_thin}, we plot $\alpha(\nu)$ in the optically thin limit for several combinations of magnetic field strength $|B|$ and electron temperature $\Theta_e$.  Thermal models are plotted with solid lines, while $\kappa=5.0$ models are plotted with dotted lines.  A dashed black line demarcates 230 GHz, the primary observing frequency of the EHT and the focus of this study.

Plasmas with both thermal and kappa eDFs have critical frequencies $\nu_\mathrm{crit} \propto \Theta_e^2 B$ where these curves turn over \citep[see][for details]{Pandya+2016}.  At fixed observing frequency, optically thin plasmas with higher temperatures or magnetic field strengths exhibit more positive spectral indices, up to a maximum value of $\alpha=1/3$ for either class of model.  On the other hand, as $\nu/\nu_\mathrm{crit} \to \infty$, $\alpha$ diverges to $-\infty$ for thermal models, but asymptotes to $\alpha_\mathrm{min} = -(\kappa-2)/2$ for kappa models, which we mark with a thin black horizontal line.  This asymptote is identical to that of a power law eDF described by $n(\gamma) \propto \gamma^{-p}$ with $p=\kappa-1$.  Any model with a uniform value of kappa throughout the simulation domain cannot exhibit a spectral index $\alpha < \alpha_\mathrm{min}$.

\subsubsection{Optically Thick}

\begin{figure*}
  \centering
  \includegraphics[width=\textwidth]{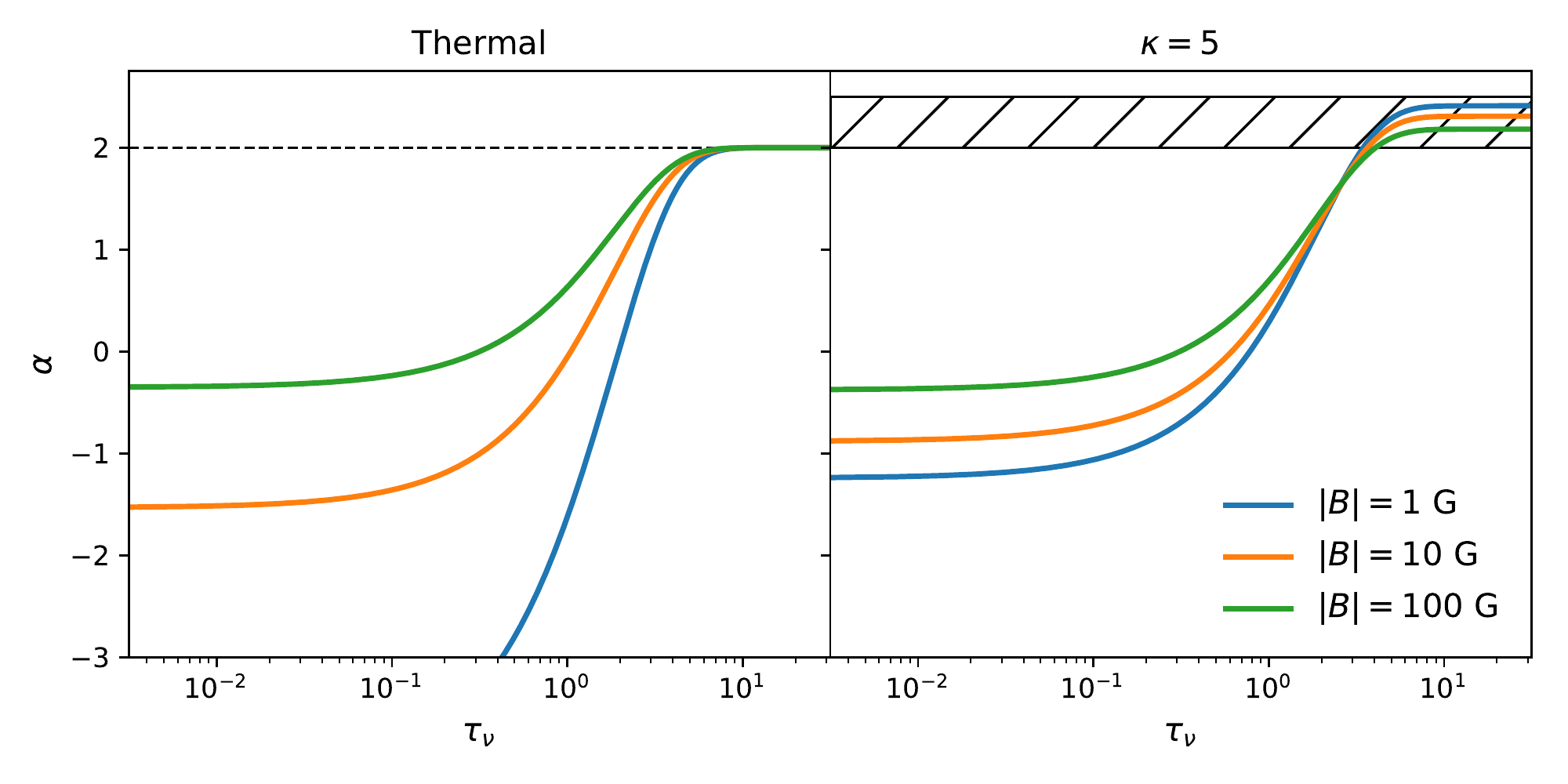}
  \caption{Spectral indices for a uniform slab of plasma as a function of optical depth for different magnetic field strengths with fixed $\Theta_e=10$ and $\nu=230$ GHz.  Increasing the optical depth increases $\alpha$ from the optically thin limit (\autoref{fig:analytic_optically_thin}) up to $\alpha_\mathrm{max}=2$ for thermal models (left; dashed black line), or $\alpha_\mathrm{max}(\kappa,\nu,B,\Theta_e) \in (2,2.5)$ for a kappa model (right; hashed region). \label{fig:analytic_spectral_index_magnetic_field}}
\end{figure*}

\begin{figure*}
  \centering
  \includegraphics[width=\textwidth]{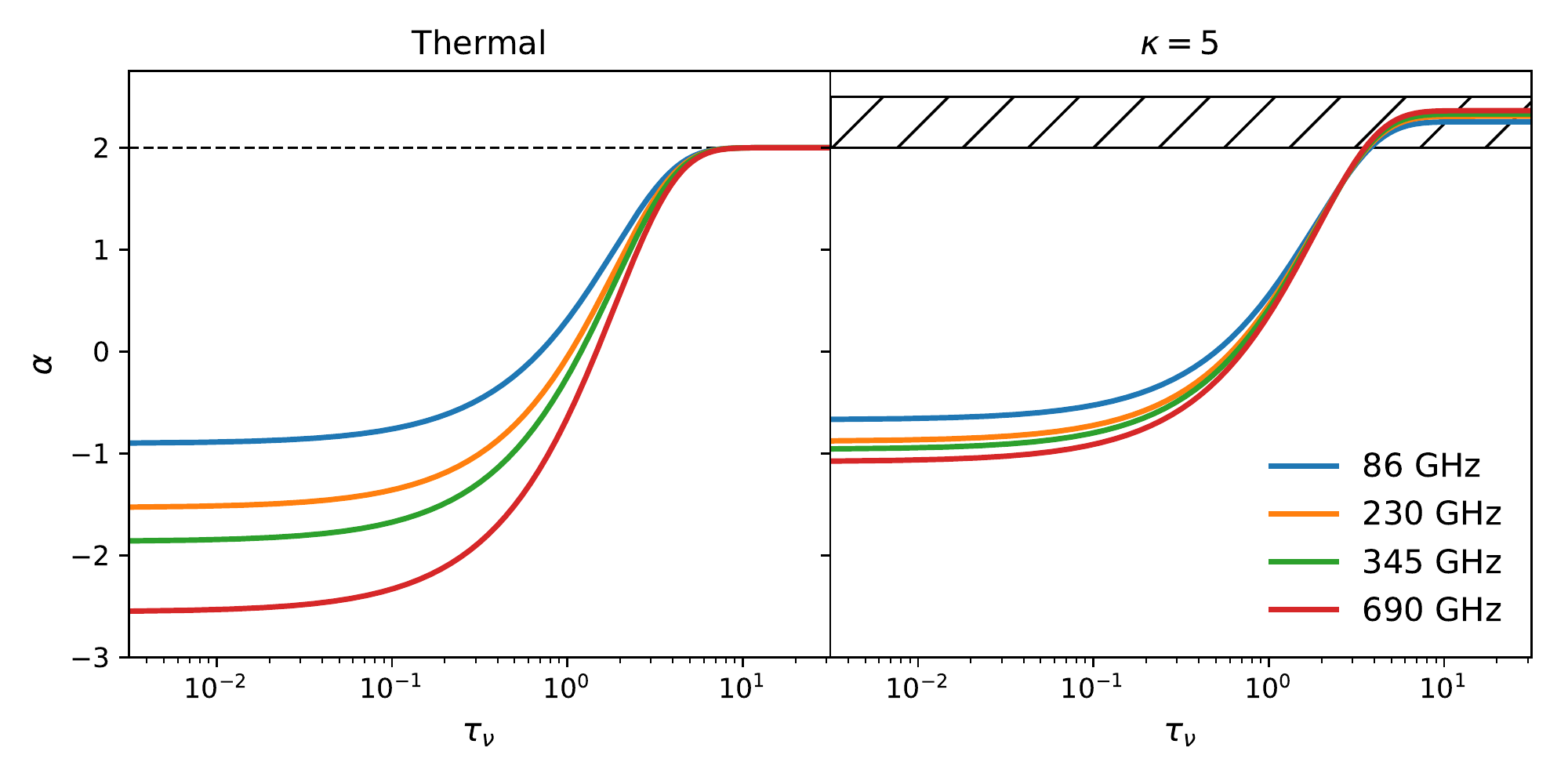}
  \caption{As \autoref{fig:analytic_spectral_index_magnetic_field}, but varying frequency instead of magnetic field strength.  Here, we fix $\Theta_e=10$ and $|B|=10 \ \mathrm{G}$.  Higher frequencies result in more negative spectral indices for small $\tau_\nu$ (see also \autoref{fig:analytic_optically_thin}), while the spectral index necessarily converges to similar values for large $\tau_\nu$.  (Note that since $\tau_\nu$ is frequency-dependent, a given $\tau_\nu$ in this plot implies a different physical column density for each frequency curve.) \label{fig:analytic_spectral_index_frequency}}
\end{figure*}

Finite optical depth always causes the spectral index of a uniform slab of material to increase.  In \autoref{fig:analytic_spectral_index_magnetic_field} and \autoref{fig:analytic_spectral_index_frequency}, we plot the spectral index as a function of optical depth, varying the magnetic field strength and the frequency respectively.  We fix $\Theta_e=10$ and $\nu=230$ GHz in \autoref{fig:analytic_spectral_index_magnetic_field} and fix $\Theta_e=10$ and $|B|=10$ G in \autoref{fig:analytic_spectral_index_frequency}.  In the optically thick limit ($\tau_\nu \to \infty$), \autoref{eqn:radiative_transfer_slab} reduces to $I_\nu = j_\nu/\alpha_\nu$, and consequently $\alpha = d\log j_\nu / d\log \nu - d\log \alpha_\nu / d\log \nu$.  For a thermal distribution, $\alpha \to 2$ (the Rayleigh-Jeans law), whereas for a kappa distribution the spectral index approaches a value between 2 and 5/2 depending on the plasma details, $\alpha_\mathrm{max}(\kappa,\nu,B,\Theta_e) \in (2,2.5)$.

In summary, the spectral index of a uniform slab of material with either a thermal or kappa eDF depends on three parameters: its magnetic field strength $|B|$, its electron temperature $\Theta_e$, and its optical depth $\tau_\nu$.  Synchrotron emitting plasma has a critical frequency $\nu_\mathrm{crit} \propto \Theta_e^2 B$, so increasing either $\Theta_e$ or $B$ will increase $\alpha$ in the optically thin limit up to a maximum value of $1/3$.  In the optically thin limit, $\alpha \in (-(\kappa-2)/2,1/3)$ for kappa eDFs, while $\alpha \in (-\infty,1/3)$ for thermal eDFs.  Optical depth can increase $\alpha$ to $\alpha_\mathrm{max}=2$ for a thermal eDF, or $\alpha_\mathrm{max}(\kappa,\nu,B,\Theta_e) \in (2,2.5)$ for a kappa eDF.

\section{Results}
\label{sec:results}

\subsection{Resolved Spectral Index Maps:  A First Glance}
\label{sec:results_resolved}

\begin{figure*}
  \centering
  \includegraphics[width=\textwidth]{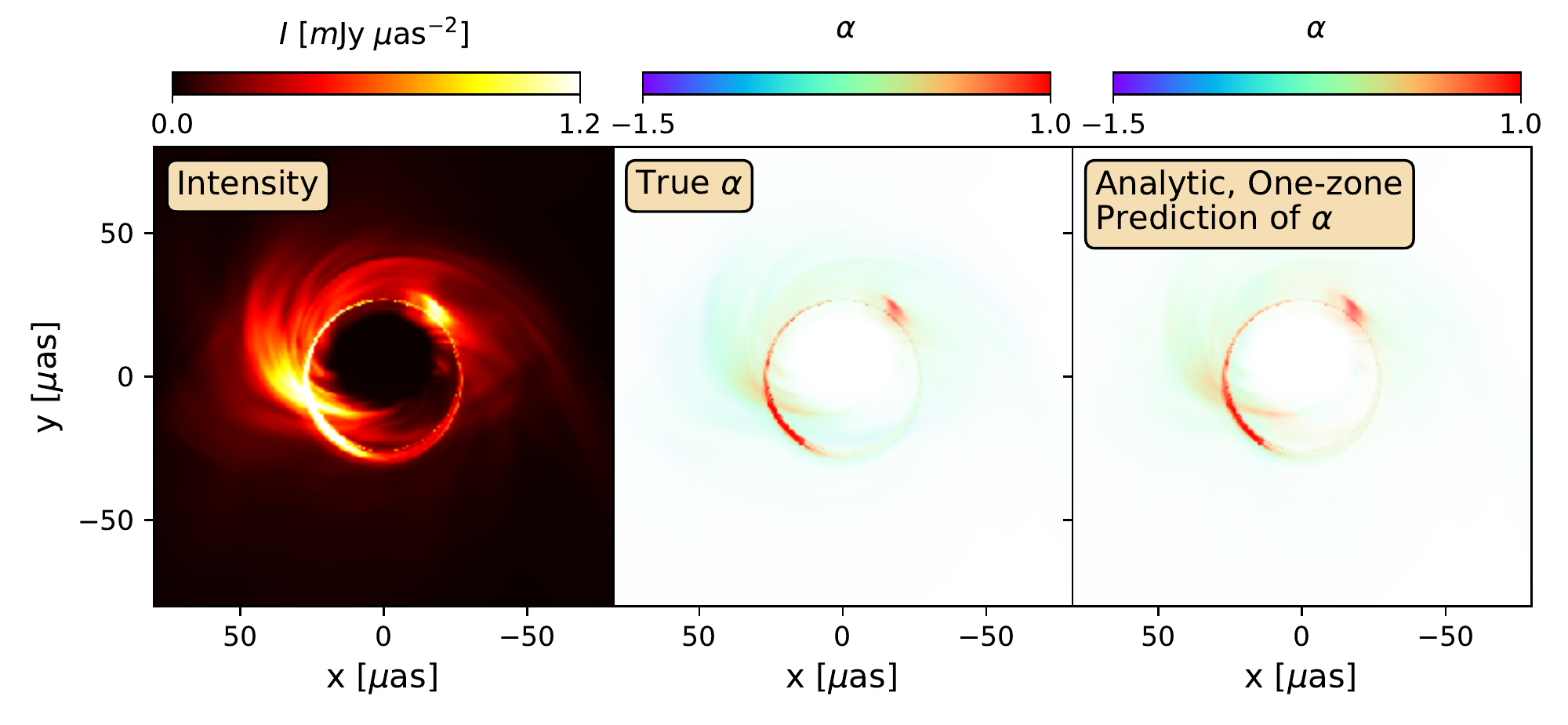} \\
  \includegraphics[width=\textwidth]{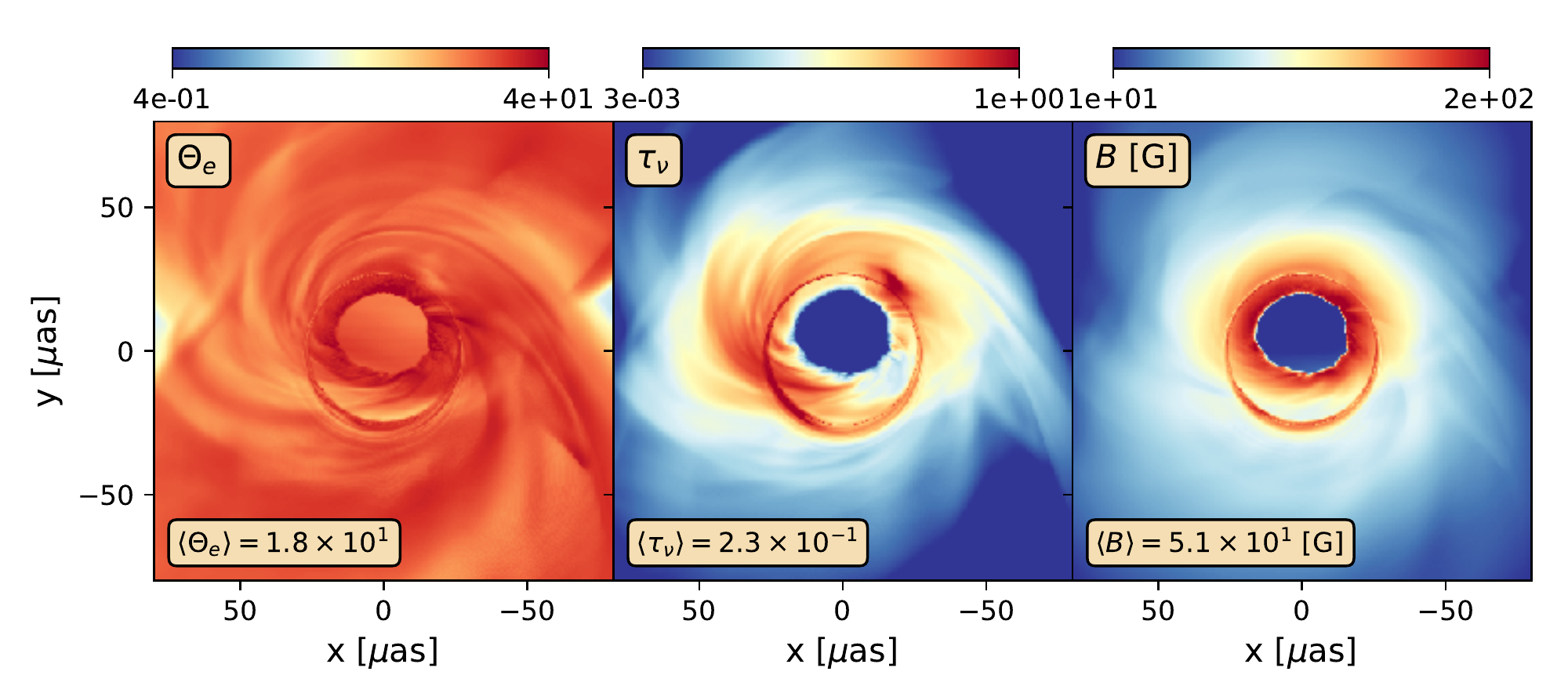} \\
  \caption{{\it Top:}  For one model of Sgr A*, an intensity map, spectral index map, and predicted spectral index map from characteristic physical quantities in each pixel between 214.1 and 228.1 GHz.  The results shown correspond to a snapshot of the MAD $a_\bullet=0$ $R_\mathrm{high}=40$ $\kappa=5$ $i=50^\circ$ model, which exhibits $\alpha_\mathrm{net}=-0.26$.  {\it Bottom:}  Pixel-wise emissivity-weighted average temperatures, optical depths, and magnetic field strengths along each geodesic.  Image-averaged intensity-weighted quantities are computed and written on the bottom of each panel.  These quantities are used to analytically predict the spatially resolved spectral index map in the top right panel, which agrees excellently with the true spectral index map computed from full ray-tracing with {\sc ipole} at two frequencies shown in the top middle panel.  Two salient features of the spectral index map can be explained by examining these physical parameters.  First, the radial decline in $\alpha$ can be explained by a corresponding decline in the magnetic field strength and optical depth.  Second, the photon ring shows up clearly on the spectral index map.  These geodesics have large optical depths due to their longer path lengths, and also because they plunge into the innermost regions of the accretion flow, with the largest magnetic field strengths.
  \label{fig:analytic_mad}}
\end{figure*}

\begin{figure*}
  \centering
  \includegraphics[width=\textwidth]{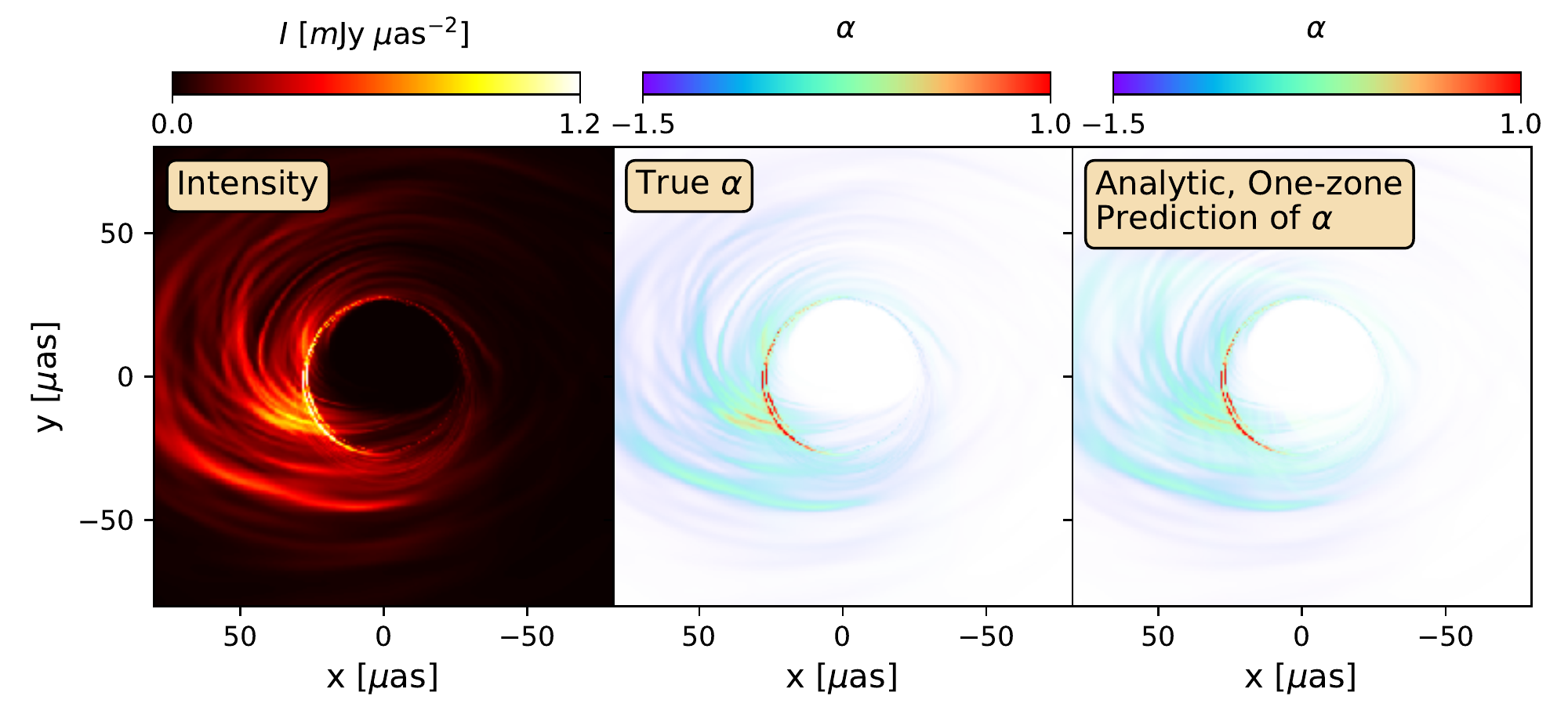} \\
  \includegraphics[width=\textwidth]{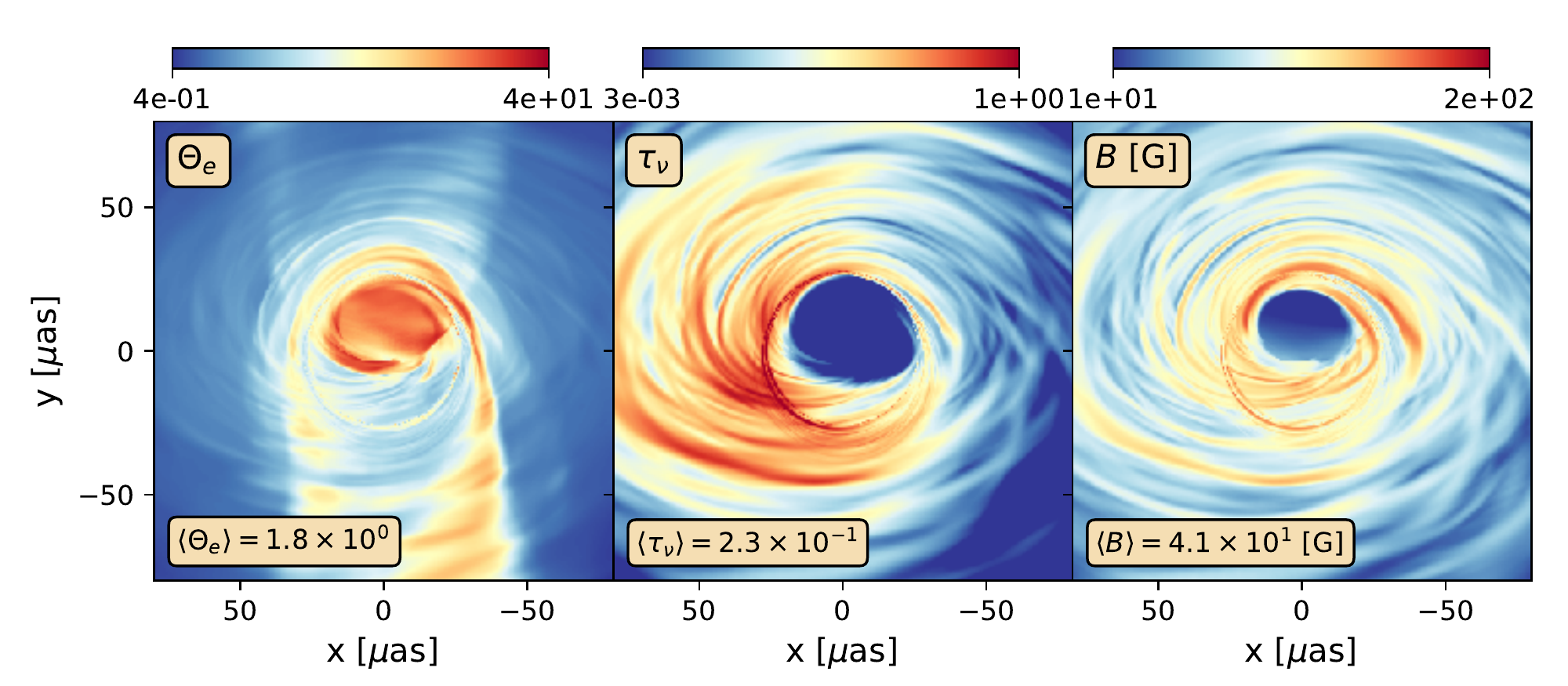} \\
  \caption{As \autoref{fig:analytic_mad}, but for the equivalent SANE snapshot (SANE $a_\bullet=0$ $R_\mathrm{high}=40$ $\kappa=5$ $i=50^\circ$).  Despite very similar optical depths and magnetic field strengths, the SANE model has a much more negative $\alpha_\mathrm{net}=-0.97$ compared to $\alpha_\mathrm{net}=-0.26$ for the MAD model.  As illustrated by the bottom left panels of these figures, this is due to the fact that the SANE disk is colder than the MAD disk by an order of magnitude.
  \label{fig:analytic_sane}}
\end{figure*}

In \autoref{fig:analytic_mad}, in the first row, we first introduce an example model's intensity map, spectral index map\footnote{Throughout this work, when plotting maps of spectral index and optical depth, we linearly scale the opacity of each pixel with the total intensity in that pixel.  We also saturate the opacity of the top 1 per cent of the emission to better visualise low-surface brightness material.}, and an analytic estimate of the spectral index map based on emission-weighted physical quantities in each pixel.  This model corresponds to a snapshot of the Sgr A* MAD $a_\bullet=0$ $R_\mathrm{high}=40$ $\kappa=5$ $i=50^\circ$ model, which has $\alpha_\mathrm{net}=-0.26$.  The spectral index becomes more negative as one moves farther from the centre of the image.  The photon ring also stands out, exhibiting a more positive spectral index than its surroundings.  

By computing emission-weighted physical quantities in each pixel, shown in the second row, we can understand these features.  Here, for a given physical quantity $\xi(\lambda)$, we assign a single emission-weighted value to each pixel $\xi \equiv \int \xi(\lambda) j_\nu(\lambda) d\lambda / \int j_\nu(\lambda) d\lambda$, where $\lambda$ is the affine parameter describing the geodesic.  Written on the bottom of each panel, we then assign a single characteristic value to each image by performing an intensity-weighted sum across all pixels, $\langle \xi \rangle \equiv \int \xi I(x,y) dx dy/ \int I(x,y) dx dy$.  Given characteristic values of $\Theta_e$, $\tau_\nu$, and $B$ in each pixel, we can then use the analysis developed in \autoref{sec:analytic} to predict $\alpha$ in each pixel.  Comparing this analytic, one-zone prediction to the true $\alpha$ computed by performing full ray-tracing with {\sc ipole} at two different frequencies, we find exceptional agreement.

Thus, we can explain the two main features in the spectral index map, its radial decline and spike at the photon ring.  Recall that $\alpha$ increases if $\nu_\mathrm{crit} \propto B\Theta_e^2$ increases.  The radial decline in $\alpha$ can be explained by a similar decline in the characterstic magnetic field strength in the emitting region.  In addition, $\alpha$ also increases if $\tau_\nu$ increases, and we see that $\tau_\nu$ indeed also decreases with radius in this image.  Finally, both $\tau_\nu$ and $B$ are largest in the photon ring, whose geodesics have the longest path lengths in the emitting region, hence the largest $\tau_\nu$, and also probe the innermost regions of the accretion flow.

In \autoref{fig:analytic_sane}, we repeat this analysis for a SANE snapshot of Sgr A* which has otherwise the same parameters (Sgr A* SANE $a_\bullet=0$ $R_\mathrm{high}=40$ $\kappa=5$ $i=50^\circ$).  This model has a much lower value of $\alpha_\mathrm{net}=-0.97$ compared to the MAD snapshot's $\alpha_\mathrm{net}=-0.26$, which we will discuss as a generic feature distinguishing the two magnetic field states in \autoref{sec:results_unresolved}.  With each model's density and magnetic field self-consistently scaled to produce an average flux of 2.4 Jy, they actually exhibit similar values of $\langle \tau_\nu \rangle$ and $\langle B \rangle$.  However, the SANE is colder than the MAD by an order of magnitude, resulting in the much lower $\alpha_\mathrm{net}$.  The tendency for MADs to be hotter than SANEs is also reported in \citet{EHT_SgrA_V}, who also find higher temperatures for larger and more prograde spins.  Note that although the jet funnel is visible in the $\Theta_e$ map (oriented vertically in projection), these regions do not contribute noticeably to the intensity, which depends also on the density and magnetic field strength.  We comment that these simulations have neglected radiative cooling, which may further reduce the temperature, and therefore also lower the spectral index.

\subsection{Changing the Electron Distribution Function}

\begin{figure*}
  \centering
  \includegraphics[width=\textwidth]{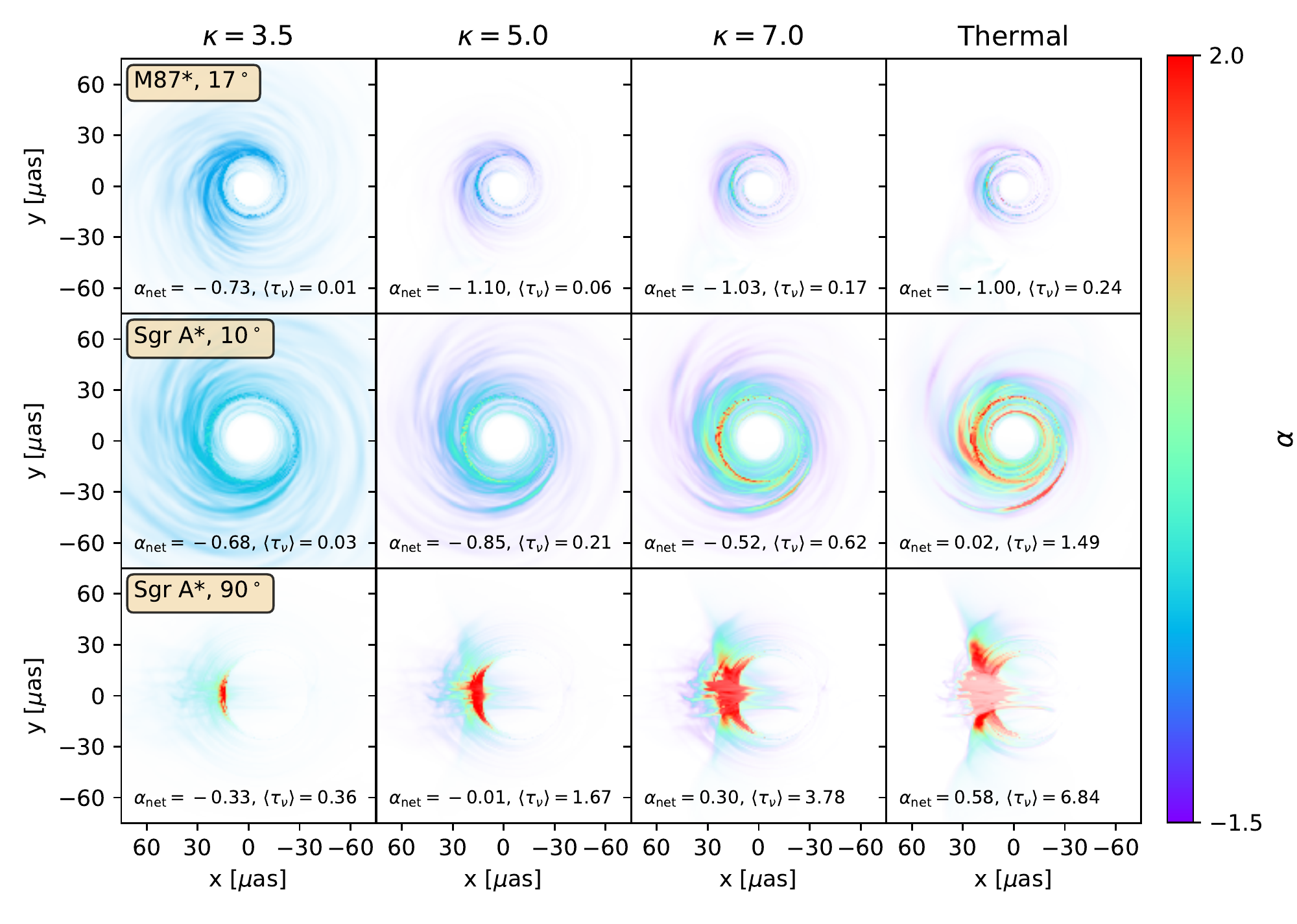} \\
  \caption{Visualisation of how the spectral index map (214.1-228.1 GHz) varies for a single snapshot as we change our choices of source, $\kappa$, and inclination angle.  The addition of a non-thermal component produces a halo of diffuse emission that grows in prominence as $\kappa$ decreases, consistent with previous studies \citep{Ozel+2000,Mao+2017}.  Models are more optically thick at larger inclinations and with smaller fractions of non-thermal electrons, therefore exhibiting more positive spectral indices.  This particular SANE is significantly optically thick for $i=90^\circ$ and $\kappa\geq 5$, leading to positive spectral indices.
  \label{fig:example_sane}}
\end{figure*}

\begin{figure*}
  \centering
  \includegraphics[width=\textwidth]{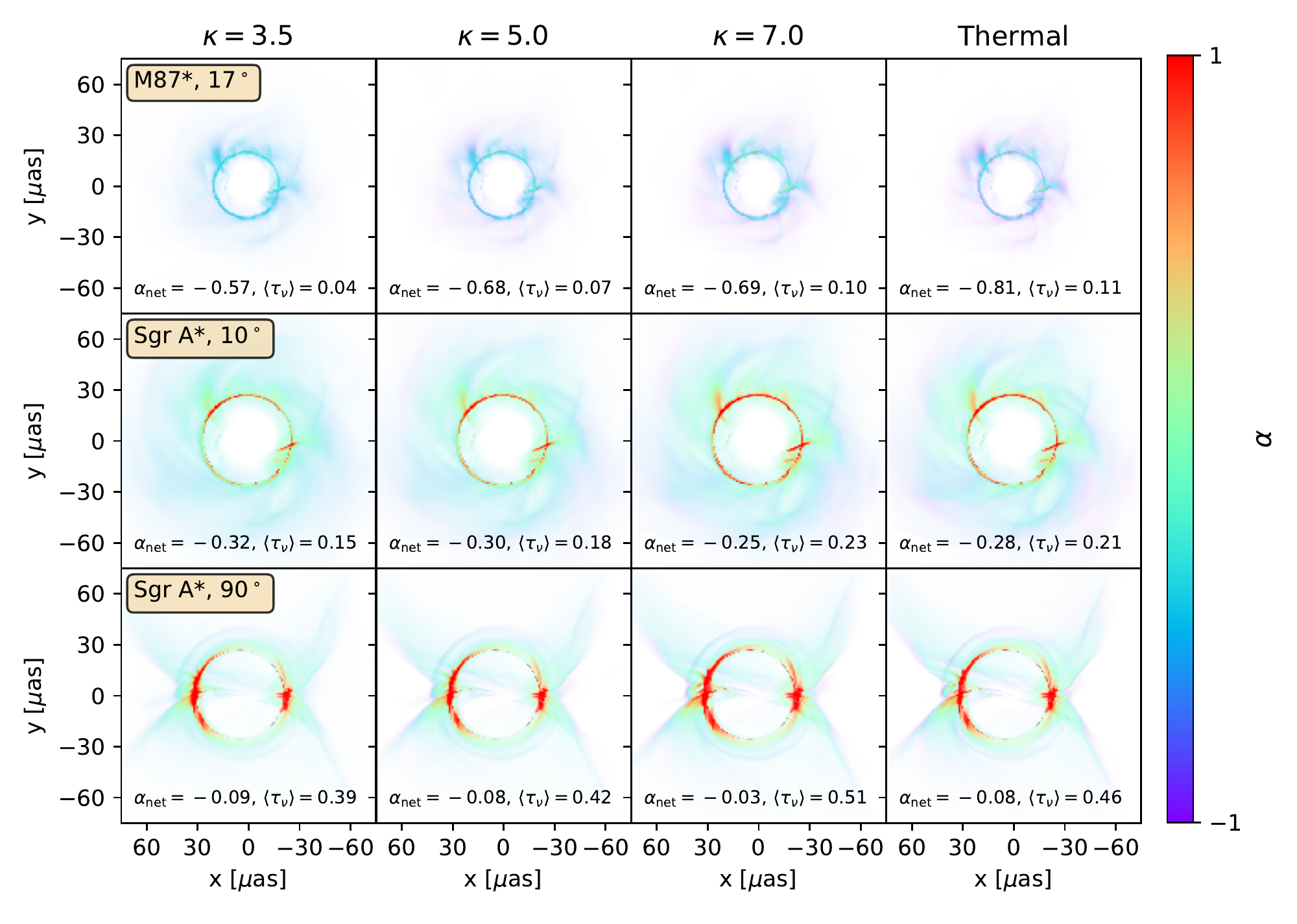} \\
  \caption{As \autoref{fig:example_mad}, but for a MAD $a_\bullet=-0.5$ $R_\mathrm{high}=160$ model, a favourable combination of parameters for M87* \citep{EHT8}.  We broadly find MAD models to be less sensitive to the choice of $\kappa$ than SANE models.
  \label{fig:example_mad}}
\end{figure*}

In \autoref{fig:example_sane}, we explore how our images and their the spectral index maps behave with respect to our choices of source, eDF, and for Sgr A*, the inclination.  The example chosen here is a SANE $a_\bullet=0.94$ $R_\mathrm{high}=40$ snapshot.  The source and viewing angle of the models in each row are written in the left-most column, while the unresolved spectral index and characteristic optical depth are written at the bottom of each panel.

Recall that smaller values of $\kappa$ correspond to a larger fraction of high-energy electrons, and that a thermal model corresponds to $\kappa \to \infty$.  Consistent with previous studies, the addition of a high-energy, non-thermal tail leads to a diffuse halo of emission which grows more prominent as $\kappa$ decreases \citep{Ozel+2000,Mao+2017}.  Since we re-normalise the accretion rate to maintain a consistent average flux for each source, the addition of more non-thermal electrons (decreasing $\kappa$) decreases the optical depth as the emitting region is spread out over a wider area.  This in turn helps decrease the spectral index as $\kappa$ decreases.  This particular SANE model exhibits a strong dependence of $\alpha_\mathrm{net}$ on the observing inclination, growing significantly optically thick for $i=90^\circ$ and $\kappa\geq 5$, resulting in flat to positive $\alpha_\mathrm{net}$.  We do not notice significant qualitative differences between our spectral index maps of M87* images and Sgr A* images, except that our Sgr A* images tend to be more optically thick and exhibit more positive spectral indices than our M87* images for the flux normalisations that we have adopted.

SANE model images have been shown to be more sensitive to electron temperature prescriptions ($R_\mathrm{high}$) than MAD models \citep{EHT5}, and the same appears to be true for the electron distribution function (eDF).  In \autoref{fig:example_mad}, we repeat this plot for a MAD $a_\bullet=-0.5$ $R_\mathrm{high}=160$ model, a favoured combination of parameters for M87* from imaging and polarimetry analysis \citep{EHT8}, and find much weaker variation.  We generally find that MAD models are less sensitive to the choice of $\kappa$ than SANEs.  
 
\subsection{Resolved Spectral Index Maps:  A Generic Radial Decline}
\label{sec:profiles}

\begin{figure*}
  \centering
  \includegraphics[width=\textwidth]{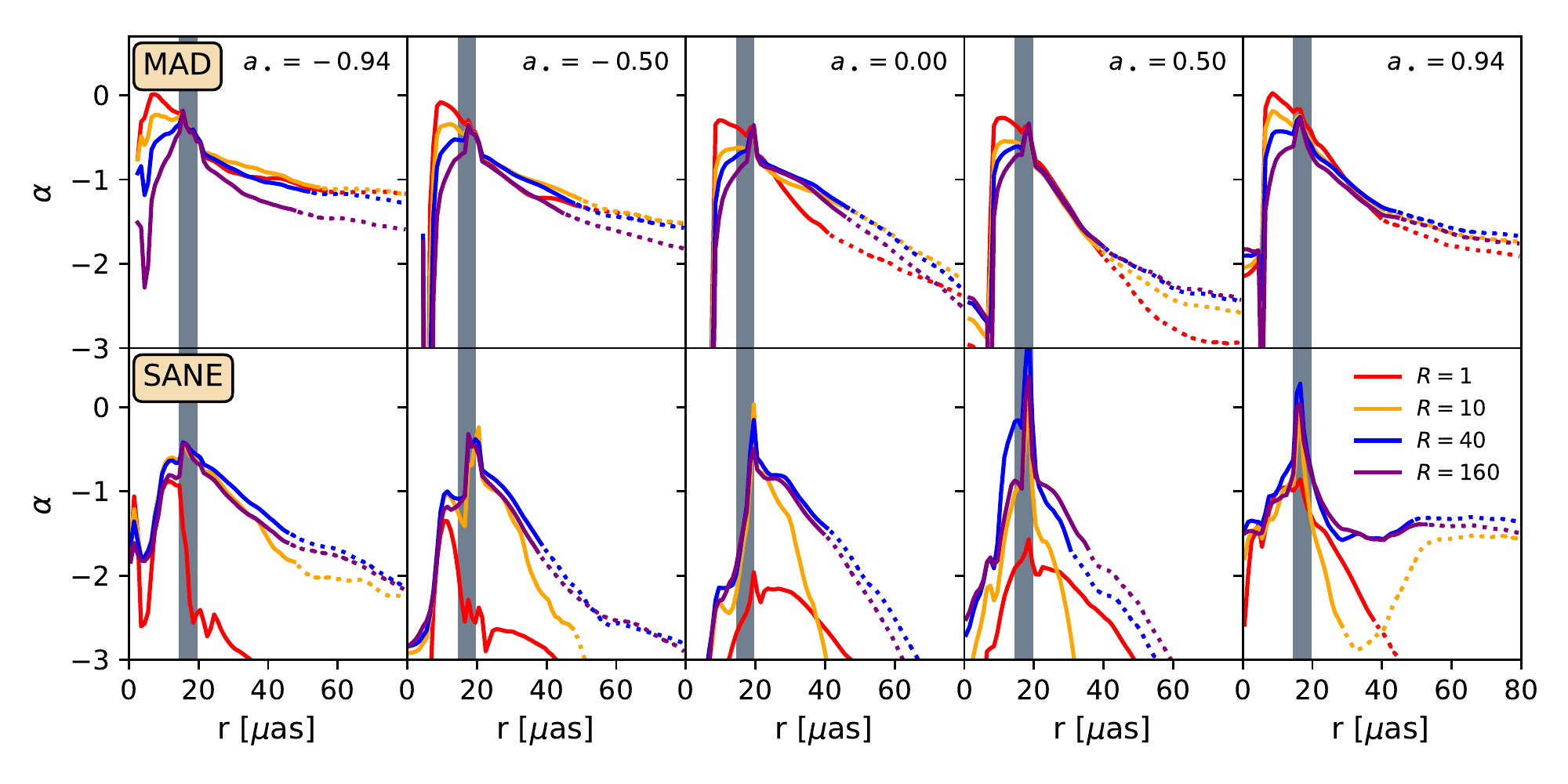} \\
  \includegraphics[width=\textwidth]{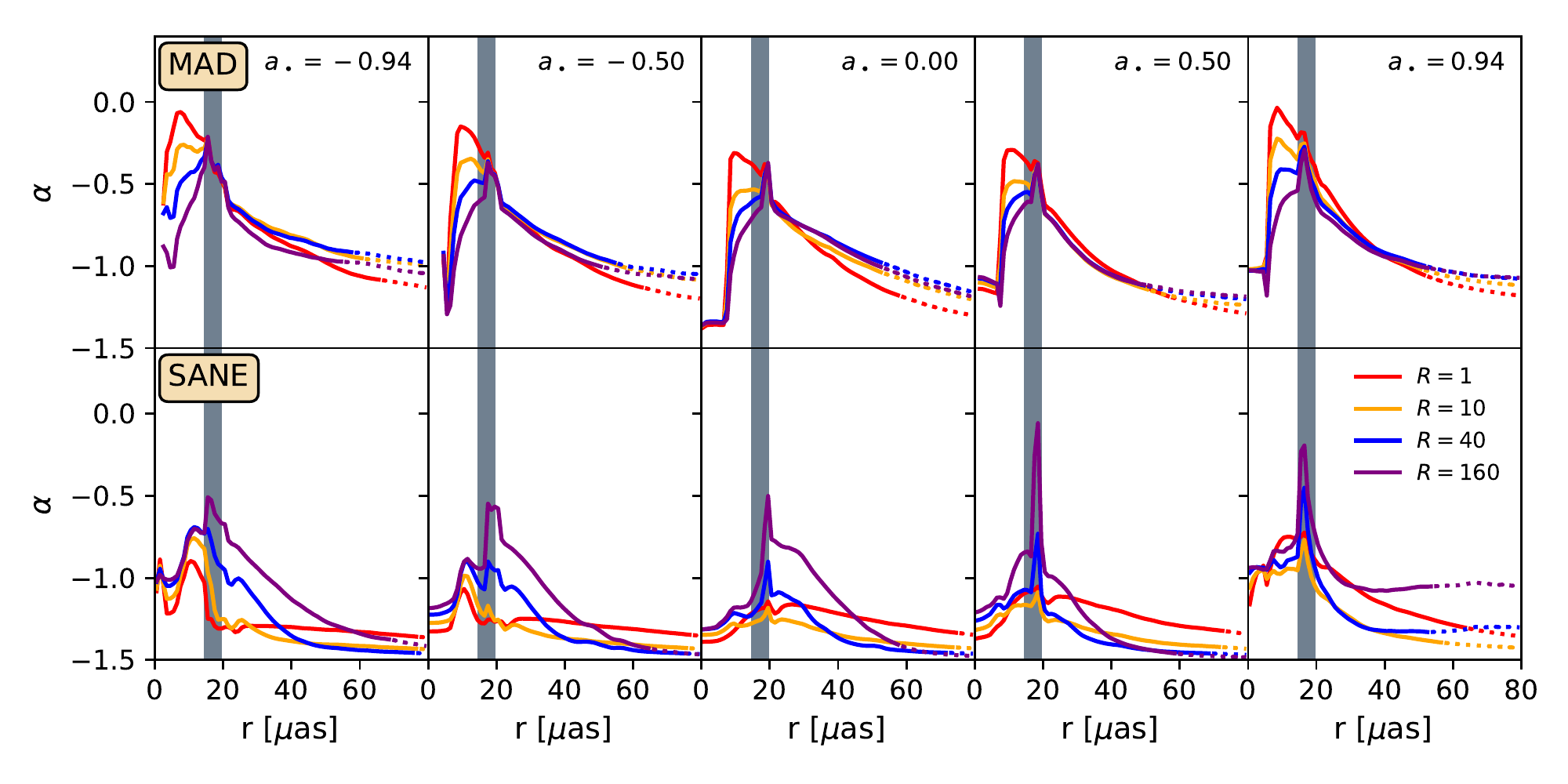}
  \caption{Azimuthally and time-averaged spectral index profiles (214.1-228.1 GHz) as a function of radius for a thermal eDF ({\it top}) and for a $\kappa=5$ eDF ({\it bottom}) for M87* models.  Note the difference in y-axis extent between the two eDF models.  Different columns correspond to different spins, different rows correspond to different magnetic field states, and the colours encode different values of $R_\mathrm{high}$.  The vertical grey band marks the radius of the photon ring, calculated analytically, which varies slightly with azimuth.  The transition from solid to dotted curves marks the radius enclosing 95 per cent of the total emission in the image.  \label{fig:profiles}}
\end{figure*}

A generic property of all resolved spectral index maps is that the spectral index grows more negative as radius increases.  In \autoref{fig:profiles}, we plot average radial profiles of the spectral index between 214.1 and 228.1 GHz for each of the M87* models.  For each model, we perform an intensity-weighted azimuthal average of the spectral index in each pixel, followed by a time-average.  The top set of panels shows results for a thermal eDF, while the lower set of panels shows results for a $\kappa=5.0$ eDF.  Note the different extent in the y-axis for the two sets of models.  Curves transition from solid to dotted at the radius enclosing 95 per cent of the total flux.

For most models, a spike in the spectral index occurs at the location of the photon ring, marked with a vertical grey band.\footnote{To compute this band, we numerically compute the minimum and maximum radius of the critical curve at a viewing angle of $17^\circ$ following the equations of \citet{Chael+2021}.}  Interior to the photon ring, MAD models with larger $R_\mathrm{high}$ have more negative spectral indices, consistent with expectations as the electron temperature decreases.  However, we caution that this region may be sensitive to our choice to zero radiative transfer coefficients within $\sigma>1$ regions.  Outside the photon ring, the spectral index universally falls off as radius increases.  This matches general expectations that the image should become smaller as the frequency increases \citep[e.g.,][]{Doeleman+2008}.  Thermal models tend to fall off more rapidly than $\kappa$ models, which are more likely to plateau at large radius.  This is because in the optically thin limit, $\kappa$ model spectral indices can never fall below $\alpha_\mathrm{min}=-(\kappa-2)/2$ (see \autoref{fig:analytic_optically_thin}).  Consequently, the spectral index in optically thin regions in the image outskirts can place a lower limit on $\kappa$.

\begin{figure*}
  \centering
  \includegraphics[width=\textwidth]{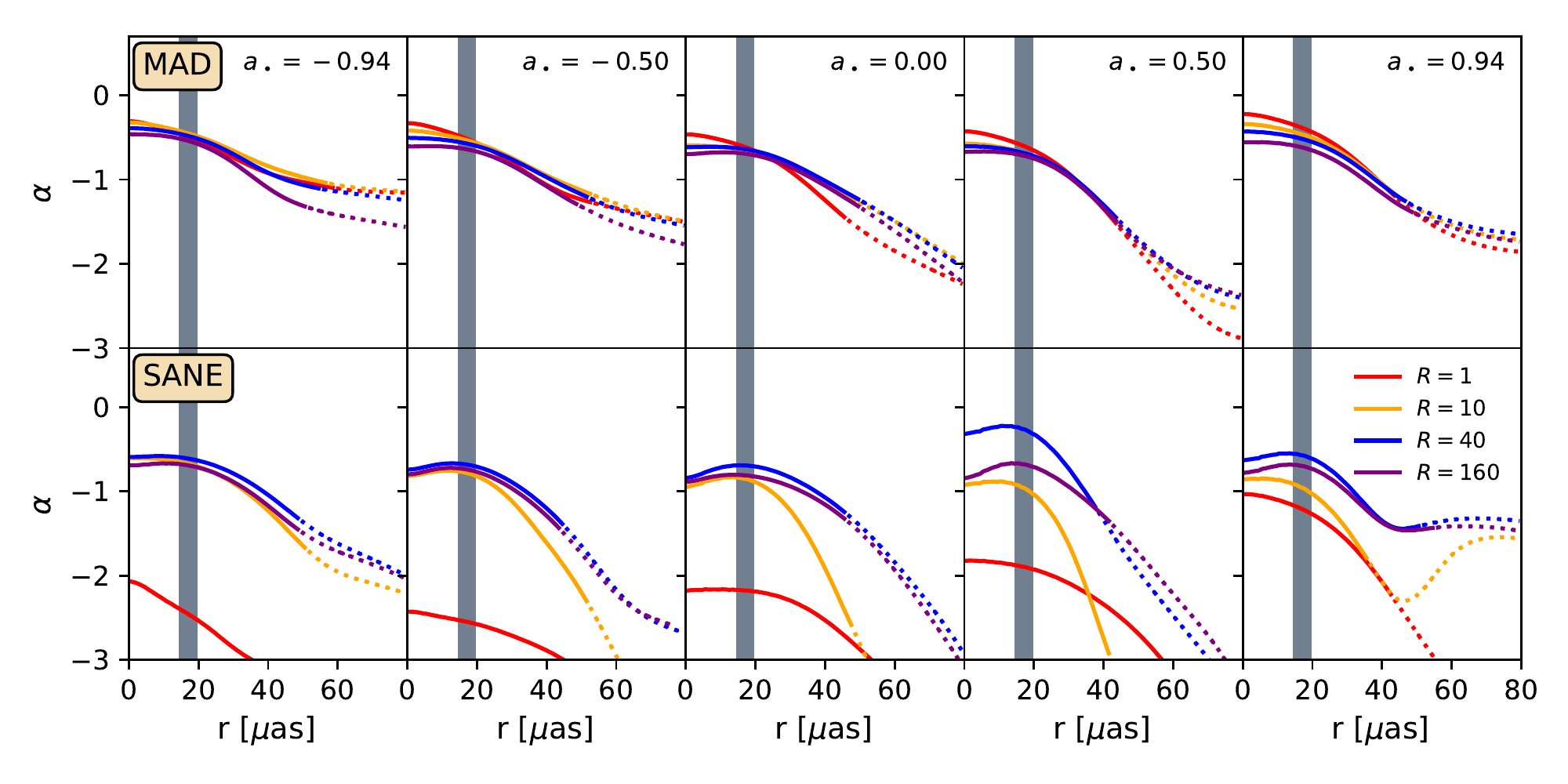} \\
  \includegraphics[width=\textwidth]{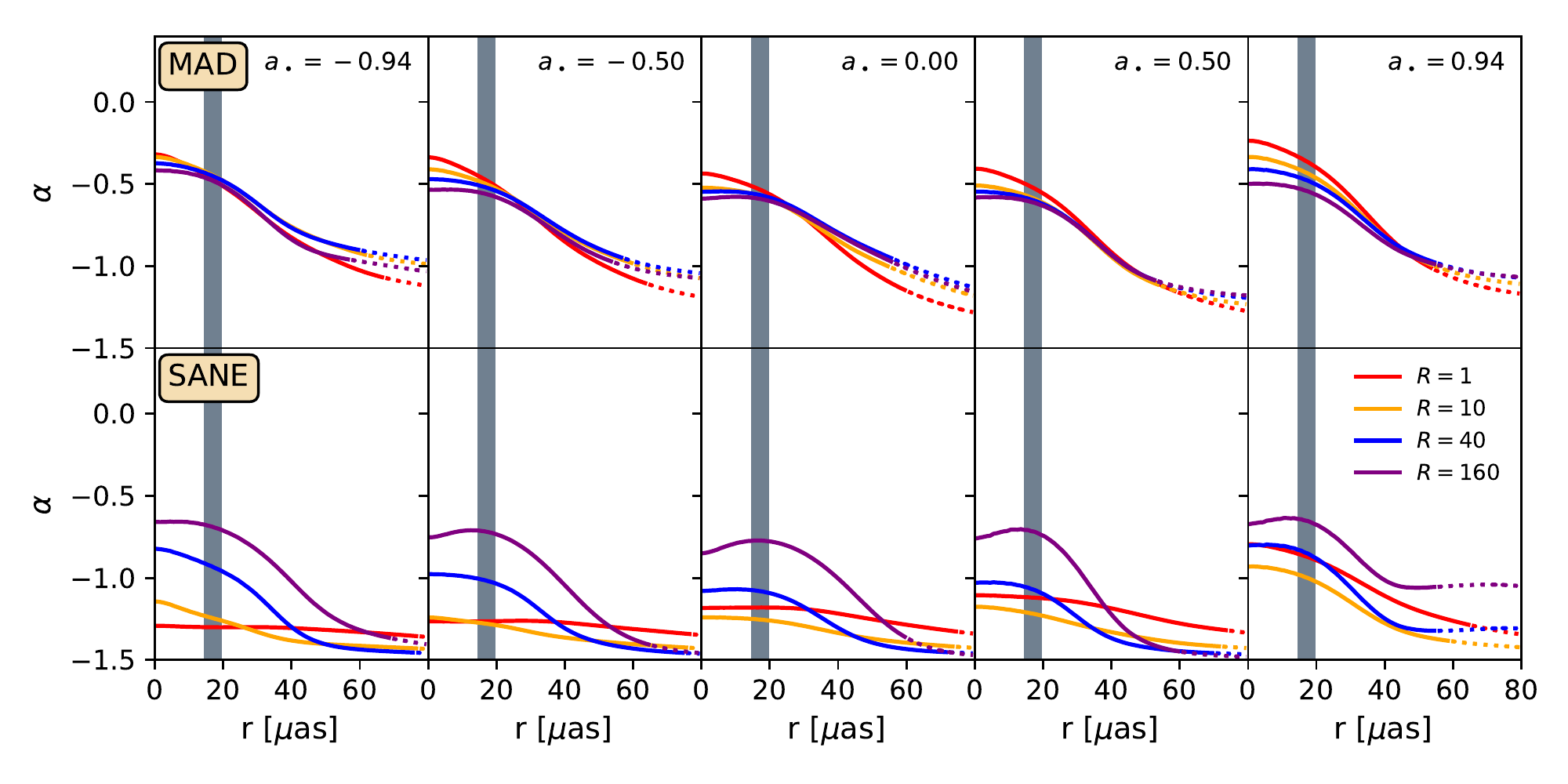}
  \caption{As \autoref{fig:profiles}, but each image is first convolved with a Gaussian with a full width at half maximum of 20 $\mu$as.  While the sharp photon ring feature is washed out, the overall radial trend is preserved. Future EHT analyses can use $d\alpha / d r$ to constrain models, for which we expect a negative value.  \label{fig:profiles_blurred}}
\end{figure*}

While \autoref{fig:profiles} assumes perfect resolution, the EHT currently only has a resolution of approximately 20 $\mu$as.  In \autoref{fig:profiles_blurred}, we repeat the analysis done to create \autoref{fig:profiles}, but first blur each image with a Gaussian with a full width at half maximum of 20 $\mu$as.  We find that the sharp photon ring feature is washed out, but the overall radial trends of these curves are preserved.  In future EHT analyses, $d\alpha / d r$ may be useful for placing constraints on models.  The shallowest gradients are exhibited by small $R_\mathrm{high}$ SANEs with a kappa eDF, which asymptote to $\alpha_\mathrm{min}$.  The steepest gradients are exhibited by thermal SANEs, due to their low temperatures.

\subsection{Unresolved Spectral Indices}
\label{sec:results_unresolved}

\begin{figure*}
  \centering
  \includegraphics[width=\textwidth]{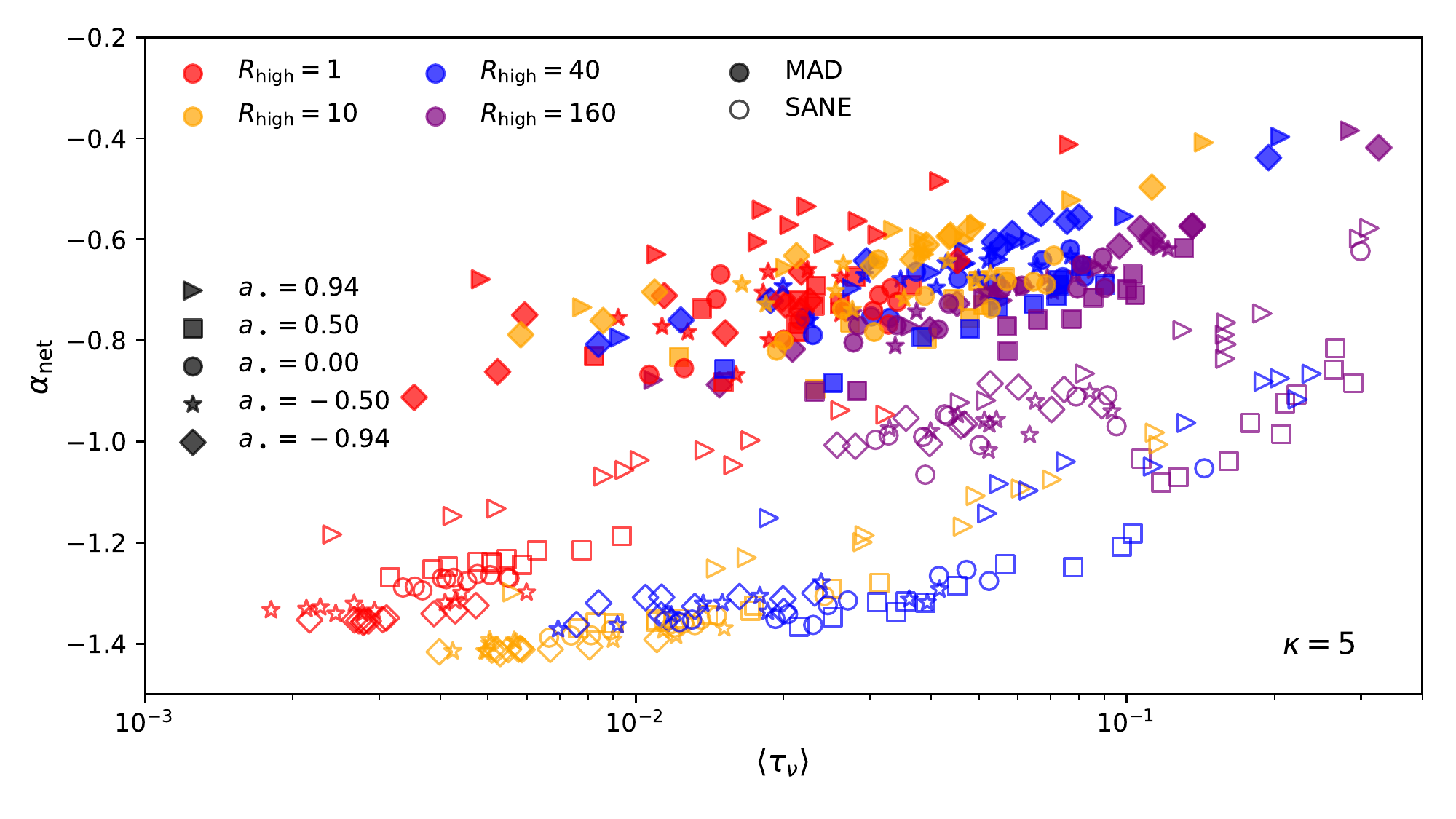}
  \caption{Unresolved spectral index ($\alpha_\mathrm{net}$, 214.1-228.1 GHz) versus optical depth ($\tau_\nu$) for M87* models with $\kappa=5$.  Different symbols correspond to different spin values, over which $\alpha$ exhibits no overall trend.  Filled symbols correspond to MADs while open symbols correspond to SANEs, each of which occupy different regions of this parameter space.  Although both classes of model span a similar range in optical depth, MAD models have more positive spectral indices than SANE models.  Different colours correspond to different values of $R_\mathrm{high}$.  For $\kappa=5$, the spectral index approaches $\alpha_\mathrm{min}=-(\kappa-2)/2=-1.5$ in the optically thin limit, towards which the SANE models asymptote} \label{fig:alpha_tau_summary_M87_k5.0}
\end{figure*}

Even lacking the spatial resolution of the EHT, the spatially unresolved spectral index, $\alpha_\mathrm{net}$, can discriminate models.  The spectral index is sensitive to uncertain plasma parameters ($\Theta_e$, eDF) as well as the magnetic field state (MAD vs. SANE), but appears relatively insensitive to spin.

We plot $\alpha_\mathrm{net}$ as a function of optical depth for $\kappa=5$ models of M87* in \autoref{fig:alpha_tau_summary_M87_k5.0}.  We focus on $\kappa=5$ for brevity, but note that the qualitative behaviour is similar for other eDFs, the main difference being the value of $\alpha_\mathrm{min}(\kappa)$.  In this plot, different colours encode different $R_\mathrm{high}$, different symbols encode different spins, and whether or not the symbol is filled encodes the magnetic field state, as indicated in the legend.

\begin{figure}
  \centering
  \includegraphics[width=0.5\textwidth]{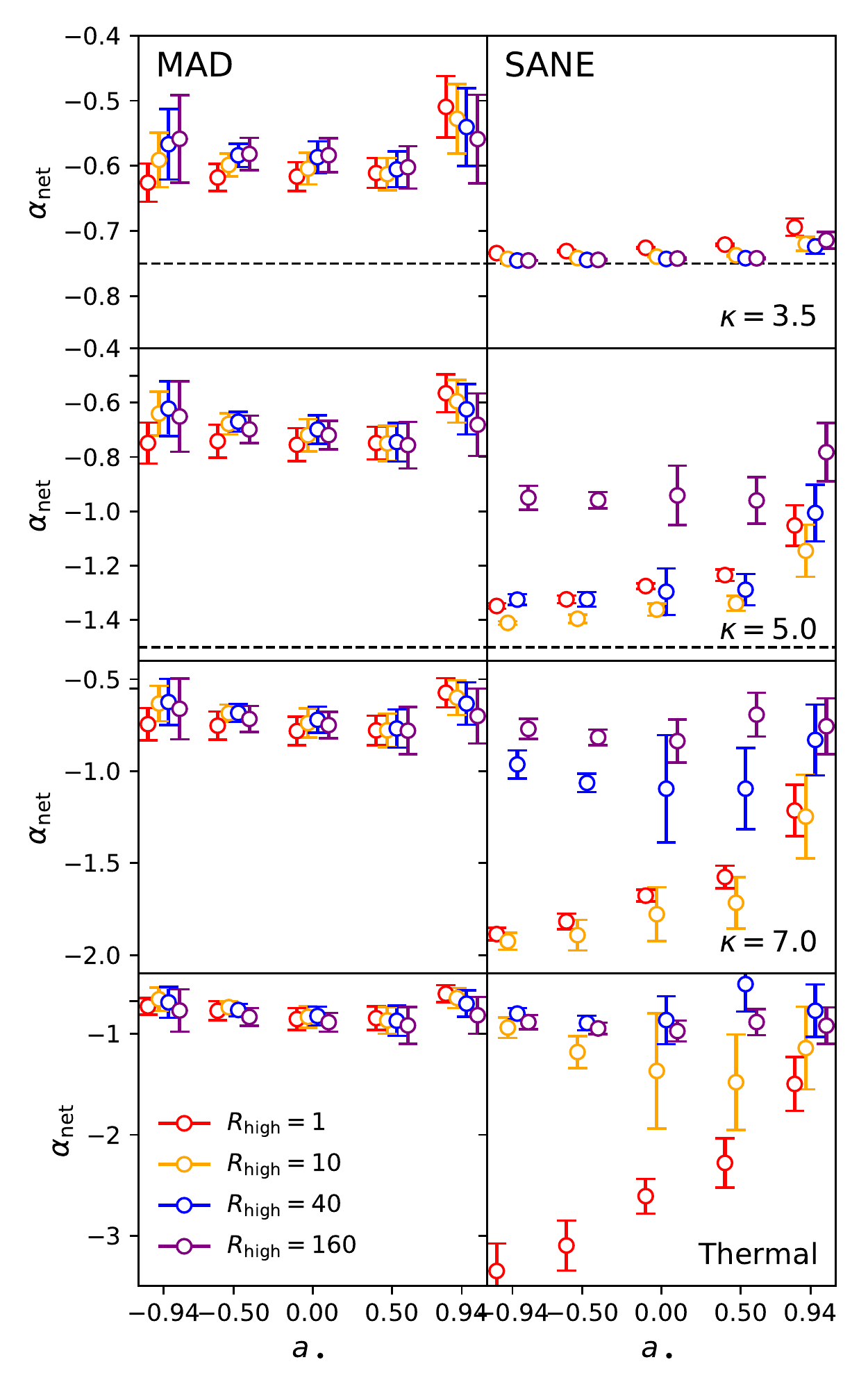}
  \caption{Unresolved spectral index ($\alpha_\mathrm{net}$, 214.1-228.1 GHz) as a function of spin for models of M87*.  The left column shows MAD models, while the right column shows SANE models.  From top to bottom, rows correspond to $\kappa=3.5$, $\kappa=5.0$, $\kappa=7.0$, and thermal models for M87*.  Different colours encode different values of $R_\mathrm{high}$, as indicated in the legend.  Where visible, the dashed black line plots $\alpha_\mathrm{min}$ for a given $\kappa$.  Error bars illustrate the standard deviation of $\alpha_\mathrm{net}$ among the 11 snapshots imaged.  Most models exhibit no discernable dependence of $\alpha_\mathrm{net}$ on $a_\bullet$.  For a given $R_\mathrm{high}$ and eDF, SANE models typically exhibit more negative $\alpha_\mathrm{net}$ than MAD models.} \label{fig:unresolved_spectral_index_spin}
\end{figure}

Our models span a wide range of both optical depth and spectral index despite all being normalised to produce the same average total intensity of 0.5 Jy.  The most apparent trend is that while the MADs (filled circles) and SANEs (open circles) span the same range of optical depth, MADs exhibit more positive spectral indices than SANEs.  This implies that MADs have larger values of $B \Theta_e^2$ in their emitting regions than SANEs.  Recall that in \autoref{fig:analytic_mad} and \autoref{fig:analytic_sane}, we found that a difference in spectral index at similar optical depth was driven by an order of magnitude difference in $\Theta_e$.  Meanwhile, there is no obvious trend in $\alpha_\mathrm{net}$ as a function of spin, which we will show more clearly in \autoref{fig:unresolved_spectral_index_spin}.  Finally, although there are trends in $\alpha_\mathrm{net}$ as a function of $R_\mathrm{high}$, these are more subtle and difficult to interpret, since the emitting region changes as $R_\mathrm{high}$ changes.  We comment that for a fixed emitting region, one would analytically expect that increasing $R_\mathrm{high}$ should decrease $\alpha_\mathrm{net}$ at fixed $\langle \tau_\nu \rangle$, since $\Theta_e$ decreases with increasing $R_\mathrm{high}$, and the MAD models are consistent with this behaviour.

\begin{figure*}
  \centering
  \includegraphics[width=\textwidth]{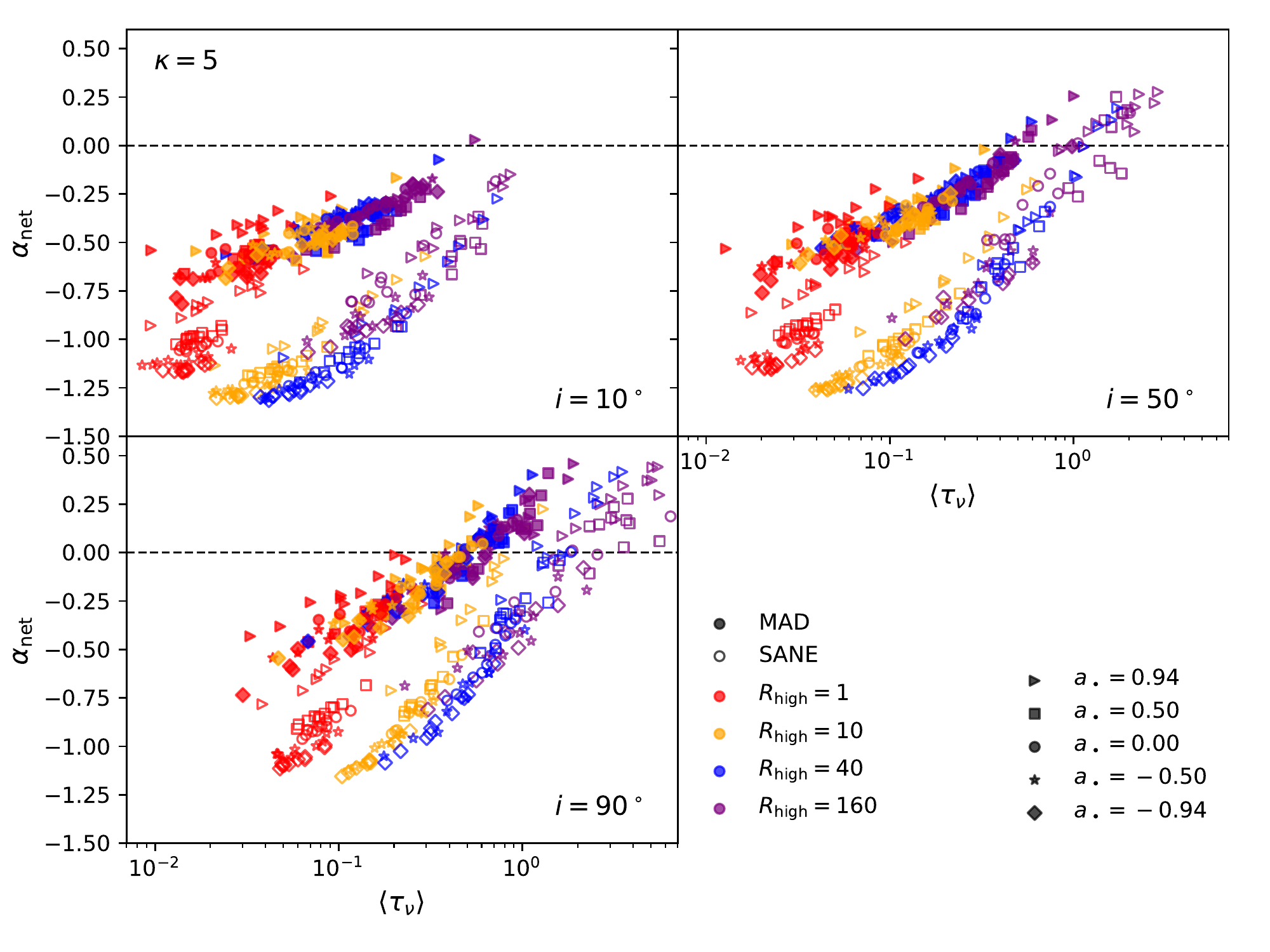}
  \caption{As \autoref{fig:alpha_tau_summary_M87_k5.0}, but for Sgr A* models with varying inclination.  Behaviour is qualitatively similar for Sgr A* compared to M87*, but models tend slightly towards larger optical depths and more positive spectral indices.  In addition, models are more optically thick and have higher spectral indices at larger inclinations. \label{fig:alpha_tau_summary_SgrA_k5.0}}
\end{figure*}

We repeat this analysis on our $\kappa=5$ models of Sgr A* in \autoref{fig:alpha_tau_summary_SgrA_k5.0}, now with a separate panel for each inclination studied.  The same qualitative trends seen for M87* persist, although Sgr A* models tend towards larger $\langle \tau_\nu \rangle$ and therefore more positive $\alpha_\mathrm{net}$ than M87* models.  As inclination increases, both $\langle \tau_\nu \rangle$ and $\alpha_\mathrm{net}$ also increase.  With larger values of $\langle \tau \rangle$, the effect of $R_\mathrm{high}$ is more pronounced for MAD models of Sgr A* than for M87*.  Our models tend to prefer more negative values of $\alpha_\mathrm{net}$ than the observed $\alpha = 0.0 \pm 0.1$, matched by only 2 MAD models with a viewing angle of $10^\circ$, but is more easily passed if the inclination is increased to $50^\circ$, by both MADs and SANEs.

To more clearly visualise differences in models as a function of spin, magnetic field state, and now also eDF, we plot for each M87* model the mean and standard deviation of $\alpha_\mathrm{net}$ for the 11 snapshots sampled in \autoref{fig:unresolved_spectral_index_spin}.  We plot MAD models on the left column and SANE models on the right column.  The different rows correspond to different eDFS, with $\kappa=3.5$, $\kappa=5.0$, $\kappa=7.0$, and thermal models represented from top to bottom.  Different colours encode different values of $R_\mathrm{high}$, as indicated in the legend.  Where visible, a dashed horizontal line marks $\alpha_{min}$, the spectral index below which a model cannot fall given a value of $\kappa$.

For the majority of the models, $\alpha_\mathrm{net}$ is generally insensitive to $a_\bullet$, which implies that $a_\bullet$ typically has little effect on $B$, $\Theta_e$, and $\tau_\nu$ of the emitting region.  The exceptions are low $R_\mathrm{high}$ SANE models with large values of $\kappa$, those SANEs which emit mostly from the disk rather than the funnel.  However, these models can already be ruled out by their overly steep spectral indices, falling below even the large-scale ALMA measurement of $\alpha \approx -1.1$ \citep{Goddi+2021}.  Some models appear to exhibit greater magnitudes of time variability than others, likely reflecting the magnitude of $d\alpha_\mathrm{net}/d\tau_\nu$ for a given set of plasma parameters.  In particular, the SANEs with $\kappa=3.5$ exhibit very little time variability, since $\alpha_\mathrm{net}$ is near the asymptotic value of -0.75 and insensitive to $\tau_\nu$.

For a fixed emitting region, one would expect that increasing $R_\mathrm{high}$ should decrease $\alpha_\mathrm{net}$, since the characteristic temperature of emitting electrons would decrease, causing $\nu_\mathrm{crit}$ to decrease.  However, recall that we scale each model with different values of $\mathcal{M}$, and the emitting region does not stay the same as $R_\mathrm{high}$ changes \citep[e.g.,][]{EHT5}.  For fixed $R_\mathrm{high}$ and eDF, MAD models usually exhibit more positive values of $\alpha_\mathrm{net}$ than SANE models.  This is because the weaker magnetic fields and lower temperatures of SANE models cause $\nu_\mathrm{crit}$ to be smaller.  For $\kappa=3.5$, all SANE models are near $\alpha_\mathrm{min}$, indicating that they are in the optically thin, high-frequency limit.

\subsection{230 vs. 345 GHz}

\begin{figure*}
  \centering
  \includegraphics[width=\textwidth]{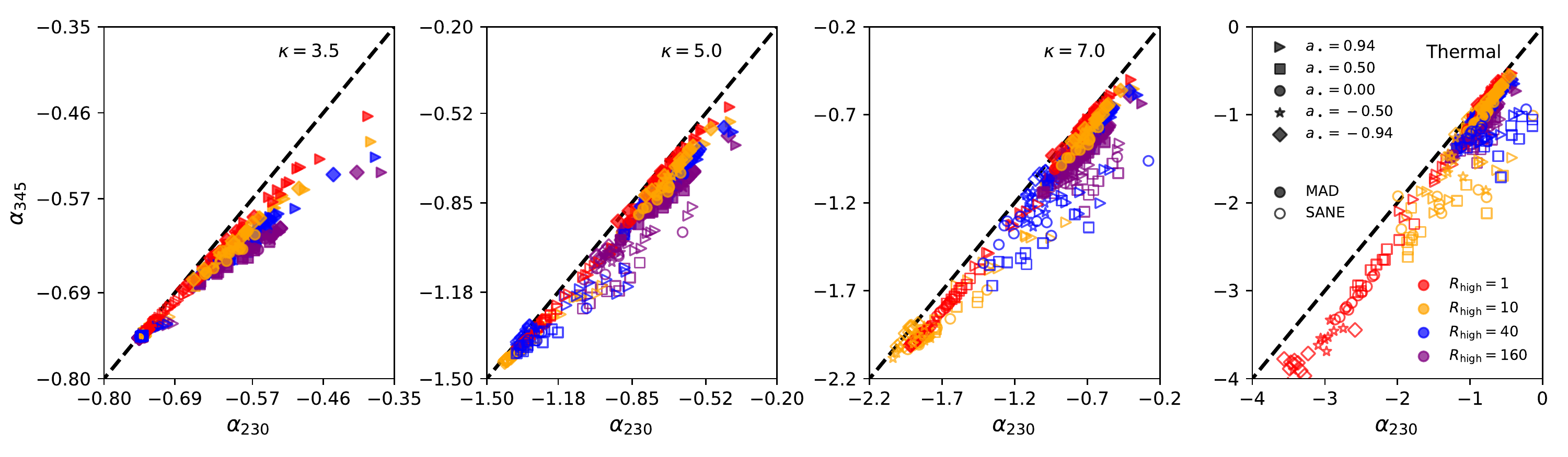}
  \caption{A comparison of spectral indices in the vicinity of 230 GHz ($\alpha_{230}$) with those in the vicinity of 345 GHz ($\alpha_{345}$).  Three different values of $\kappa$ are shown, 3.5, 5.0, and 7.0 from left to right.  As anticipated by \autoref{fig:analytic_spectral_index_frequency}, increasing the observing frequency typically decreases the spectral index for two reasons: (i) the optical depth decreases with frequency, and (ii) $\nu/\nu_\mathrm{crit}$ increases with frequency, which will decrease the spectral index even in the optically thin limit (see \autoref{fig:analytic_optically_thin}). \label{fig:230_345}}
\end{figure*}

There are currently plans to expand the observing frequency of the EHT from 230 GHz to 345 GHz \citep{Doeleman+2019,Raymond+2021}.  In addition to providing better spatial resolution, 345 GHz images will allow us to explore frequency-dependent signatures in two complementary bands.  Here, we explore the degree to which $\alpha$ depends on the observing frequency by comparing $\alpha_{230}$ between 214.1 and 228.1 GHz and $\alpha_{345}$ between 340 and 350 GHz.  This probes the second derivative of the flux with respect to frequency, while the spectral index probes the first.  Since 230 GHz is near the maximum frequency of emission from synchrotron \citep[e.g.,][]{Gravity+2020,EHTMWL+2021}, one could expect the second derivative to be significant.

We plot the spatially unresolved $\alpha_{230}$ versus $\alpha_{345}$ for our M87* models in \autoref{fig:230_345}, where from left to right panels correspond to $\kappa$ of 3.5, 5.0, and 7.0, following the plotting scheme introduced in \autoref{fig:alpha_tau_summary_M87_k5.0}.  Note the different axis limits in each panel.  A dashed diagonal line marks $\alpha_{230} = \alpha_{345}$, which almost all models fall below.  That is, increasing the observing frequency decreases the observed spectral index, as anticipated from \autoref{fig:analytic_spectral_index_frequency}.  This occurs for two reasons.  First, for a uniform parcel of gas, increasing the observing frequency decreases the optical depth, which then decreases the spectral index.  Second, even in the optically thin regime, the spectral index decreases as $\nu/\nu_\mathrm{crit}$ increases (see \autoref{fig:analytic_optically_thin}).  The frequency-dependence of $\alpha$ may help reconcile the flat spectral index observed by \citet{Kim+2018} for M87* between 22 and 128 GHz with the more negative spectral indices our models produce at 230 GHz.

\section{Discussion}
\label{sec:discussion}

\subsection{Signatures in Visibility Amplitudes}

\begin{figure*}
  \centering
  \includegraphics[width=\textwidth]{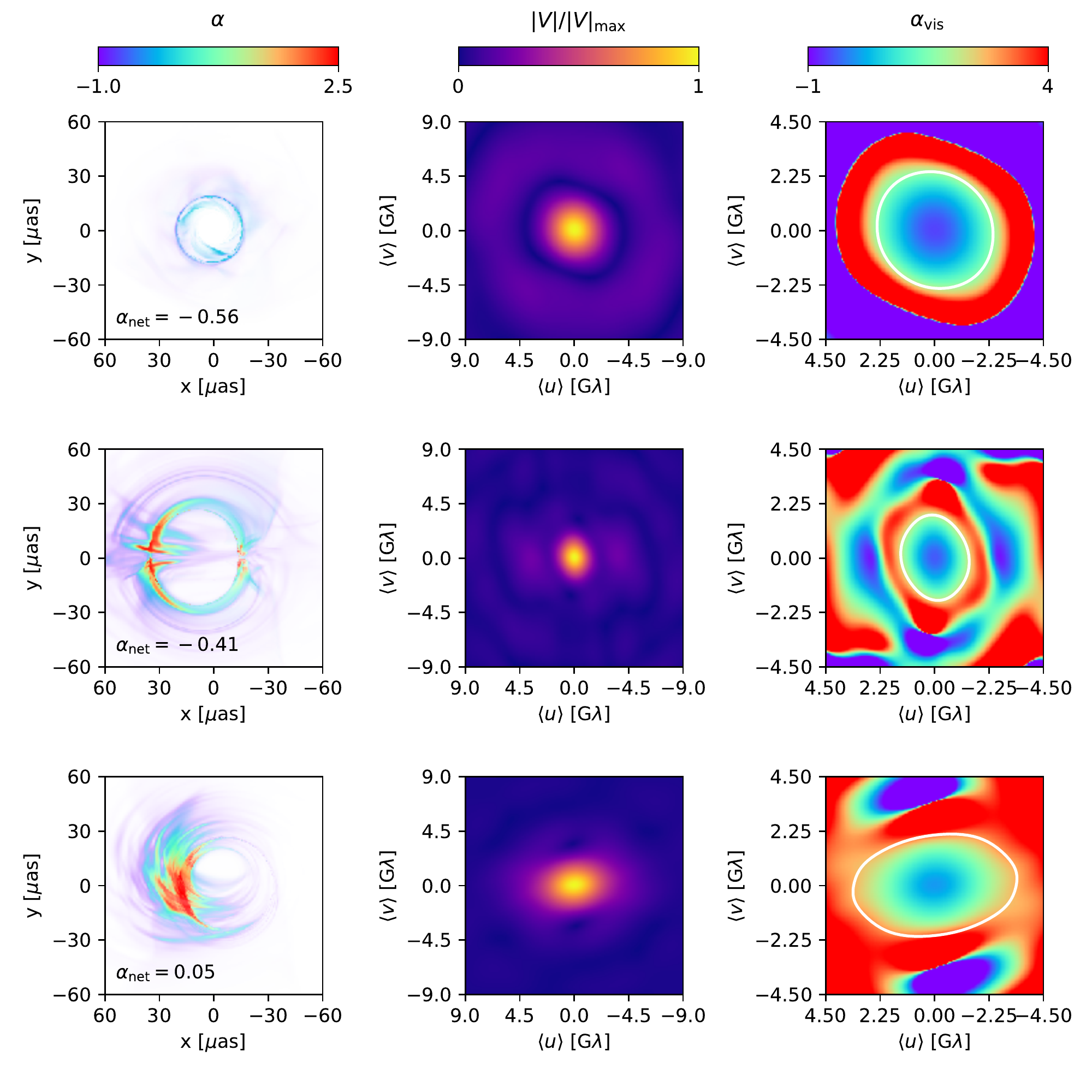} 
  \caption{For three models, we plot their spectral index maps between 214.1 and 228.1 GHz (left), visibility amplitudes (centre) and spectral indices in visibility space (right).  Models correspond to M87* MAD $a_\bullet=-0.94$ $R_\mathrm{high}=1$ thermal $i=17^\circ$, Sgr A* SANE $a_\bullet=-0.94$ $R_\mathrm{high}=160$ $\kappa=5$ $i=90^\circ$, and Sgr A* SANE $a_\bullet=+0.94$ $R_\mathrm{high}=40$ $\kappa=7$ $i=50^\circ$.  Here, $\langle u \rangle$ and $\langle v \rangle$ correspond to the {\it average} values of $u$ and $v$ between 214.1 and 228.1 GHz.  That is, we keep the baselines fixed, which would otherwise move as a function of frequency.  In the rightmost column, we zoom in towards the central region, marking with a white curve where the visibility amplitude drops below 1/3 of its maximum.  Within this boundary, $\alpha_\mathrm{vis}$ increases with radius, which may be observable on the shortest EHT baselines \citep[such as SMT-LMT;][]{EHT1}.  This occurs because as frequency increases, the image becomes smaller, and therefore its main peak grows larger in Fourier space. \label{fig:visibilities}}
\end{figure*}

To avoid uncertainties in image reconstruction, it may be helpful to search for signatures of spectral evolution in visibility amplitudes.  Here, we define {\it for a fixed baseline} $\alpha_\mathrm{vis} = d\log |V|/ d\log \nu$, where $|V|$ is the visibility amplitude.  Fixing a baseline is distinct from fixing a point in the (u,v) plane, since the (u,v) distance sampled by a baseline is proportional to the observing frequency.

We find that for short baselines, those before the first null, $\alpha_\mathrm{vis}$ generically declines as a function of (u,v) distance.  We plot three examples in \autoref{fig:visibilities}, where a resolved spectral index map is plotted in the first column, visibility amplitudes are plotted in the second column, and $\alpha_\mathrm{vis}$ is plotted in the third column.  From top to bottom, these plots feature models (M87* MAD $a_\bullet=-0.94$ $R_\mathrm{high}=1$ thermal $i=17^\circ$), (Sgr A* SANE $a_\bullet=-0.94$ $R_\mathrm{high}=160$ $\kappa=5$ $i=90^\circ$), and (Sgr A* SANE $a_\bullet=+0.94$ $R_\mathrm{high}=40$ $\kappa=7$ $i=50^\circ$).  Note that the 2017 EHT observations extended to a maximum of $\approx 8 \; \mathrm{G}\lambda$. In the rightmost column, a white curve marks where the visibility amplitude drops below 1/3 of its maximum value.

The guiding intuition behind this trend is that $\alpha$ generically declines as a function of radius in real space (see \autoref{fig:profiles}), such that the most diffuse emission has the most negative $\alpha$.  As (u,v) distance increases, this diffuse emission is resolved out, resulting in a more positive $\alpha_\mathrm{vis}$.  Beyond the central peak in Fourier space (roughly past the white contour), $\alpha_\mathrm{vis}$ becomes muddled, since $|V|$ is sensitive to changes in the intrinsic source structure as a function of frequency.  Equivalently, Sgr A* and M87* should appear smaller with increasing frequency, which causes the main peak of its Fourier transform to grow larger.  New imaging algorithms have been developed to simultaneously capture total intensity, spectral index, and even spectral curvature maps to account for these subtleties \citep{Chael+2022}.

\subsection{A Spectral Index of Polarization?}
\label{sec:polarization}

Here, we briefly explore how fractional linear and circular polarization may change as a function of frequency.  We define a spectral index of linear polarization $\alpha_p = d\log m / d\log \nu$ and a spectral index of circular polarization $\alpha_v = d\log |v| / d\log \nu$, where $m$ and $v$ are the spatially unresolved fractional linear and circular polarizations respectively.  Faraday rotation is the dominant depolarization mechanism in these models \citep{Moscibrodzka+2017,Ricarte+2020,EHT8}, and thus the general expectation is that the linear polarization should increase as the Faraday rotation depth decreases as $\nu^{-2}$.  Meanwhile, the circular polarization fraction can exhibit many different power law dependencies on frequency depending on its origin, as discussed in detail in \citet{Wardle&Homan2003}.  The circular polarization fraction should exhibit $v\propto \nu^{-1/2}$ for intrinsic emission, $v \propto \nu^{-3}$ for Faraday conversion mediated by a twist in the magnetic field, and $v \propto \nu^{-5}$ for Faraday conversion mediated by Faraday rotation.  These different pathways for generating circular polarization in the context of EHT sources are discussed at length in \citet{Ricarte+2021}.

\begin{figure}
  \centering
  \includegraphics[width=0.5\textwidth]{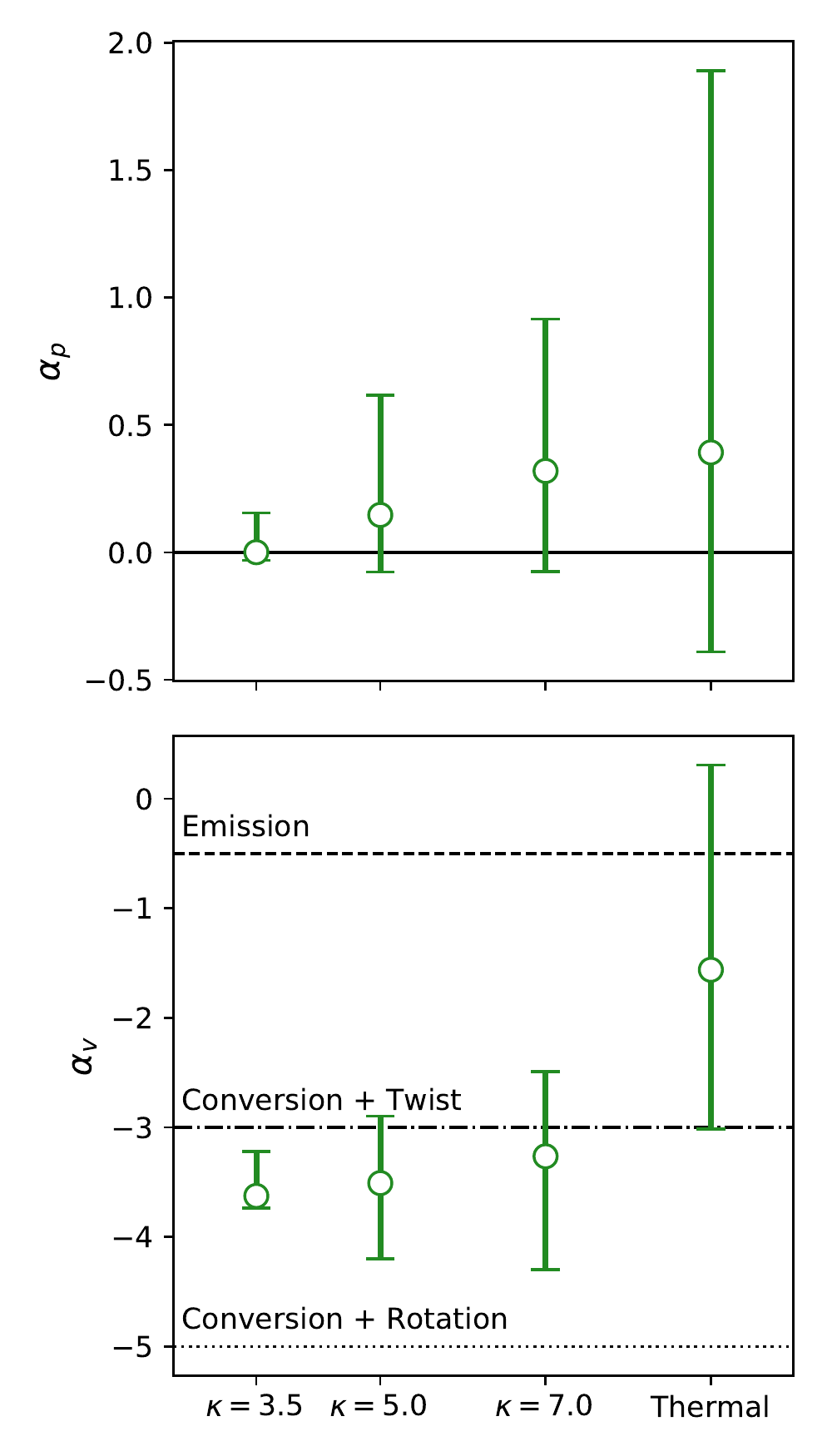}
  \caption{Distributions of spatially unresolved spectral indices of fractional linear polarization, $\alpha_p$ (top) and fractional circular polarization, $\alpha_v$ (bottom) between 214.1 and 350.0 GHz for models of M87*.  Circles mark the median of these distributions, while errorbars span between the 16th and 84th percentiles.  $\alpha_p$ is usually positive, consistent with decreasing Faraday depolarization at higher frequencies.  Meanwhile, $\alpha_v$ is more negative for kappa models than thermal models, which may suggest a change in the relative importances of pathways generating circular polarization.  Analytic expectations for $\alpha_v$ from a single zone of emission are plotted with horizontal lines.  \label{fig:polarization_spectral_index}}
\end{figure}

Despite these analytic expectations, we find a surprising amount of scatter in $\alpha_p$ and $\alpha_v$ for a given eDF, with no clear patterns with respect to $R_\mathrm{high}$, spin, or magnetic field state (\autoref{fig:polarization_spectral_index}).  We calculate $\alpha_p$ and $\alpha_v$ between 214.1 GHz and 350.0 GHz for our M87* models.  We include scatter plots of these quantities in Appendix \ref{sec:polarization_appendix}.  We do, however, notice trends as a function of eDF (\autoref{fig:polarization_spectral_index}).  For each eDF, circles demarcate the 50th percentile of the distribution of these polarization spectral indices, with error bars extending to the 16th and 84th percentiles.  As shown in the top panel, $\alpha_p$ is usually (but not always) positive.  Exceptions to the rule can arise from either optical depth effects (different fractional polarizations at different optical depths) or the spatial complexity of the Faraday rotating structure\footnote{The Faraday rotator is not spatially uniform and is coincident with the emitting region.  Since radiation from different emitting regions experiences different amounts of Faraday rotation on the way to the observer, polarization which is Faraday rotated to cancel at one frequency may instead sum at another frequency.} \citep{Ricarte+2020,EHT8}.  Meanwhile, as plotted in the bottom panel, we find that $\alpha_v$ is more negative for kappa models than for thermal models.  As horizontal lines, we plot analytic expectations if the circular polarization arises {\it solely} from a single circular polarization generation pathway.  Using thermal models, \citet{Ricarte+2020} showed that all three pathways are important for producing circular polarization in models of M87*.  Differences in $\alpha_v$ may hint at a change in the dominant production pathways as a function of eDF towards Faraday conversion for kappa models instead of intrinsic emission for thermal models.

\section{Conclusions}
\label{sec:conclusion}

The EHT has recently produced images of both M87* and Sgr A*, resolved on event horizon scales.  This allows us to place indirect constraints on important properties of the space-time and its accretion flow, such as its spin, magnetic field state, and temperature.  Spectral index information has not been included in analysis so far.  On the theoretical side, predictions require at least doubling the image library across the frequency axis and are sensitive to the eDF, but are otherwise straightforward.  On the observational side, a greater bandwidth is necessary to produce reliable maps in the presence of calibration errors and image reconstruction artifacts.  However, spatially unresolved measurements can already place meaningful constraints on our models and help break degeneracies between parameters such as density, magnetic field strength, and temperature.

In this work, by performing GRRT calculations on a broad set of GRMHD simulations, we have studied the 230 GHz spectral index $\alpha$ of models of both M87* and Sgr A*.  We consider both thermal and non-thermal (kappa) models of the eDF, and explore different electron temperature prescriptions as well as viewing angles for Sgr A*.  Our results are summarised as follows.

\begin{itemize}
    \item We first establish a theoretical foundation to interpret the spectral index by studying a uniform slab of emitting plasma.  Notably, the spectral index decreases as $\nu/\nu_\mathrm{crit}$ increases, where $\nu_\mathrm{crit} \propto B \Theta_e^2$.  This behaviour drives most of the trends we see in our models.
    \item Consequently, $\alpha$ tends to be larger for MAD models than for SANE ones, since $B \Theta_e^2$ is higher for MADs than for SANEs.  Since all models are scaled to produce the same 230 GHz flux, their differences in spectral index demonstrates the ability for spectral index to break model degeneracies.
    \item For resolved images, these models generically predict that $\alpha$ should decrease with radius, due to decreasing temperature and magnetic field strength, as well as decreasing optical depth.
    \item If resolvable by future instruments, the photon ring should exhibit a sharp peak in $\alpha$, joining a chorus of polarimetric signatures:  predictable shifts in the EVPA pattern \citep{Himwich+2020}, linear depolarization \citep{Jimenez-Rosales+2021}, and an inversion in the sign of circular polarization \citep{Moscibrodzka+2021,Ricarte+2021}.
    \item Even the spatially unresolved spectral index $\alpha_\mathrm{net}$ varies substantially between models and can in principle be used to distinguish them.  However, the exact value of $\alpha_\mathrm{net}$ is sensitive to the most uncertain parameters of GRRT---the electron temperature and eDF prescriptions---and should be interpreted cautiously.
    \item We show both analytically and numerically that $\alpha$ generically decreases with increasing frequency, due both to decreasing optical depth and increasing $\nu/\nu_\mathrm{crit}$.
\end{itemize}

While most EHT studies only consider thermal eDFs, we have explored the importance of the eDF in our work by studying kappa distributions, characterised by a thermal core and a non-thermal tail \citep{Vasyliunas1968}.  These distributions are motivated both observationally by the solar wind \citep{Decker&Krimigis2003,Pierrard&Lazar2010} and theoretically by detailed particle-in-cell simulations \citep{Kunz+2016}. Characterising this non-thermal population has been a major uncertainty in our accretion flow models.  We demonstrate a sensitivity of our images and their spectral indices to the non-thermal eDF slope, and constraining this slope will help constrain electron acceleration mechanisms.  As a first step, we have only considered models in which the eDF varies globally, whereas a more physical model may allow $\kappa$ to vary with local plasma parameters \citep[e.g.,][]{Davelaar+2018,Fromm+2022,Cruz-Osorio+2022,Scepi+2022}.  We expect spectral index maps to be sensitive to these details, which would be useful to explore in future works.  In addition, since the spectral index is sensitive to temperature, radiative cooling may also play a role in setting the spectral index of M87* in particular.  Naively, including radiative cooling should decrease the spectral index by decreasing the temperature, but the net effect is unclear since our simulations are rescaled to match the total flux.  The effect of radiative cooling is likely to only be understood satisfactorily with a different set of GRMHD simulations including such physics.

Future observations with the EHT and ngEHT with large enough bandwidth to produce spectral index maps will enable us to constrain the magnetic field strength, temperature and electron distribution of these models.  At present, EHT imaging combined with multi-wavelength constraints favour a spinning MAD model for M87* \citep{EHT8}, and tentatively prefer the same for Sgr A*, although no model passes all constraints \citep{EHT_SgrA_V}.  Sgr A* exhibits an unresolved spectral index of $\alpha \approx 0.0 \pm 0.1$ \citep[]{Goddi+2021,Wielgus+2022}, and our models usually predict more negative values than this (see \autoref{fig:alpha_tau_summary_SgrA_k5.0}).  This may result in an indirect constraint on inclination that should be considered together with other constraints on Sgr A* in future work.  Meanwhile, M87*'s spectral index is much more uncertain due to the unknown contribution of the large-scale jet flux, but spectral index maps will be made possible with future EHT data sets.  For M87*, our MAD models of M87* tend to predict spectral indices around $-0.7$ regardless of spin, $R_\mathrm{high}$, and eDF, but SANE models produce a larger range that tends towards more negative values.  SANE models of M87* with larger values of $R_\mathrm{high}$ tend towards more positive values of $\alpha_\mathrm{net}$.  With resolved spectral index maps, the EHT will be able to indirectly constrain radial gradients of temperature, magnetic field strength, and density.  In resolved maps, our SANE models exhibit shallower gradients with radius if there is a larger non-thermal eDF population.  Under the assumption of a universal value of $\kappa$, models are also bounded from below by $\alpha_\mathrm{min}(\kappa)$ at all radii, and thus the minimum $\alpha$ observed places a lower limit on $\kappa$.  This information will contribute to an abundance of data about our EHT sources, which will allow us to break degeneracies and narrow even further the allowed models for these accretion flows.

\section{Acknowledgements}

We thank our anonymous referee and Alejandra Jimenez-Rosales for their thorough readings and helpful comments throughout the manuscript.  We further acknowledge many fruitful and invigorating discussions with our EHT colleagues Michael Johnson, Avery Broderick, Daniel Palumbo, Zachary Gelles, and Elizabeth Himwich, George Wong, and Maciek Wielgus.  This material is based upon work supported by the National Science Foundation under Grant No. OISE 1743747.  This research was made possible through the support of grants from the Gordon and Betty Moore Foundation and the John Templeton Foundation. The opinions expressed in this publication are those of the author(s) and do not necessarily reflect the views of the Moore or Templeton Foundations.  This work used the Frontera and Longhorn resources at Texas Advanced Computing Center through allocation AST20023.

\section{Data Availability}

The data underlying this article will be shared on reasonable request to the corresponding author.

\bibliography{ms}

\appendix

\section{Complete Scatter Plots of Polarization Spectral Indices}
\label{sec:polarization_appendix}

\begin{figure*}
  \centering
  \begin{tabular}{cc}
  \includegraphics[width=0.5\textwidth]{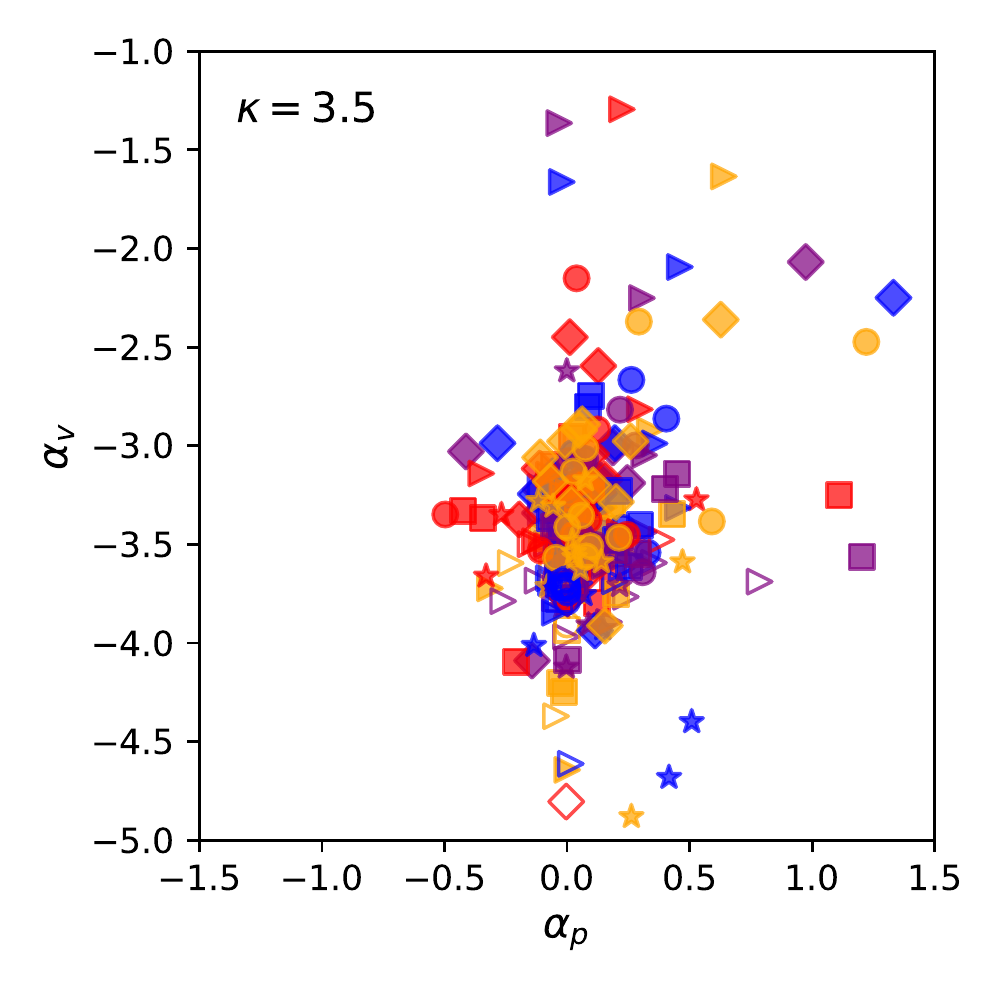} & 
  \includegraphics[width=0.5\textwidth]{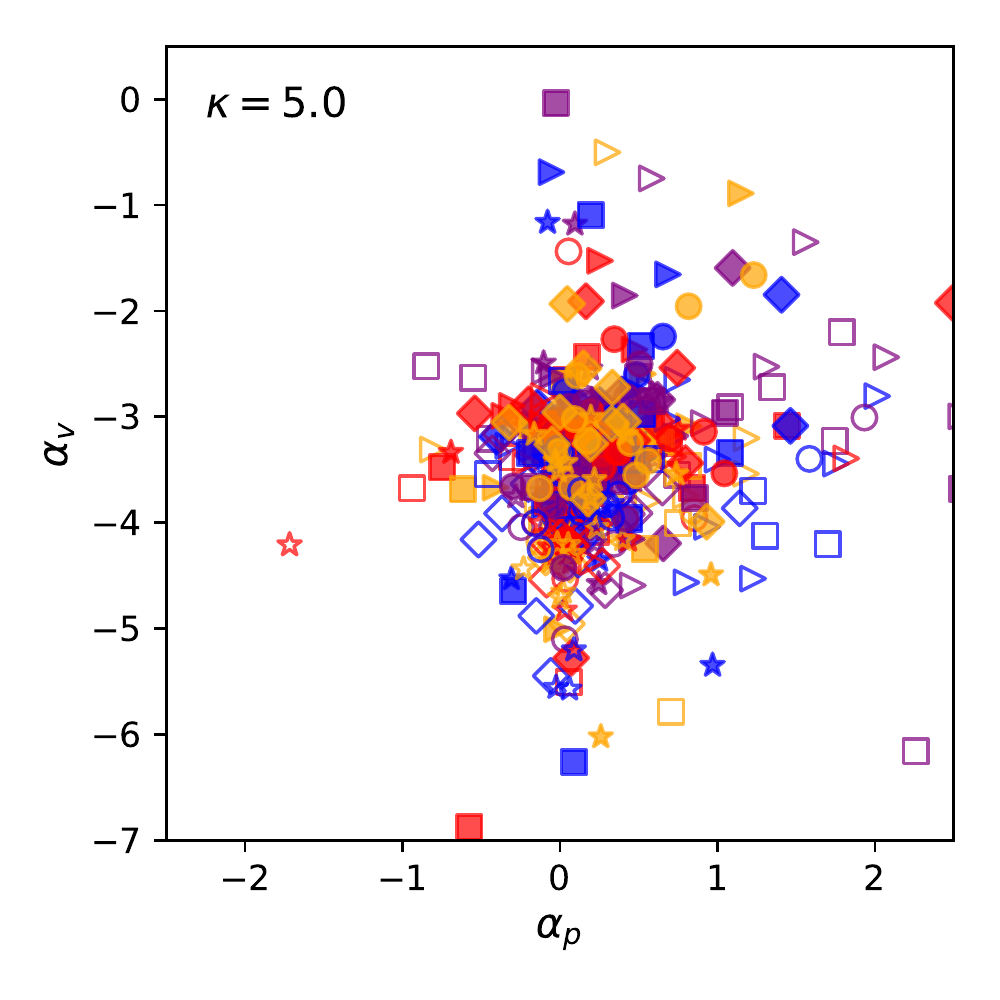} \\
  \includegraphics[width=0.5\textwidth]{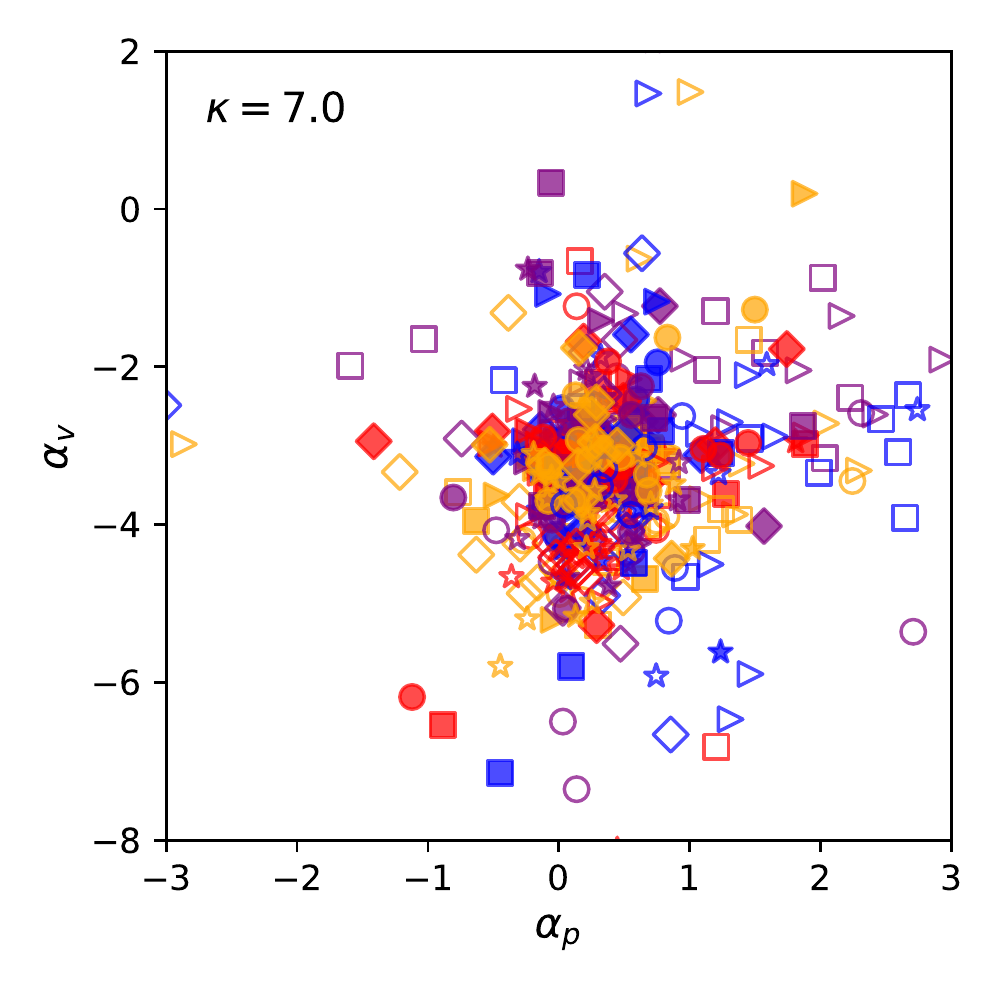} & 
  \includegraphics[width=0.5\textwidth]{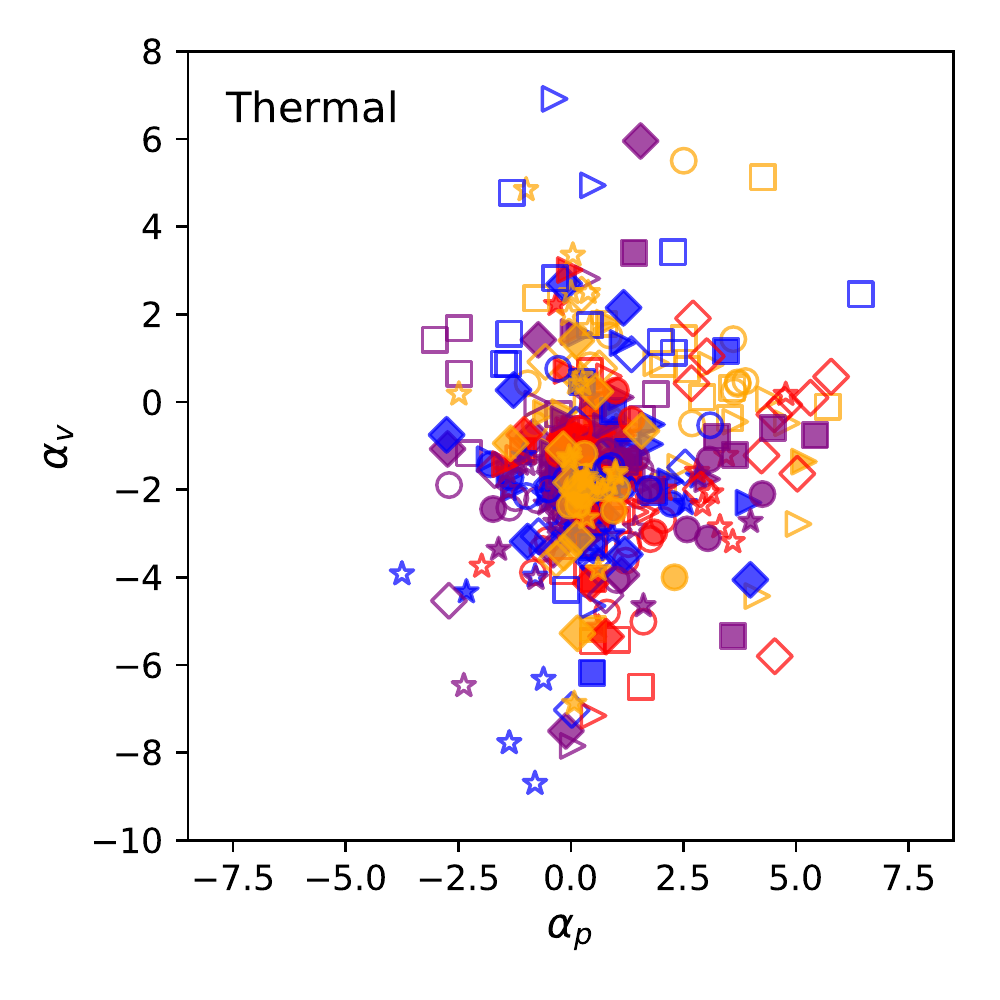} \\  
  \end{tabular}
  \caption{Polarization spectral indices between 214.1 and 350.0 GHz for our M87* models, using the plotting scheme introduced in \autoref{fig:alpha_tau_summary_M87_k5.0}.  Each panel plots a different eDF.  Note the different extents of each of the axes for each eDF.  We do not find any trend as a function of spin, magnetic field state, or $R_\mathrm{high}$, but the models of each eDF cluster in different regions, as shown in \autoref{fig:polarization_spectral_index}.  \label{fig:polarization_spectral_index_scatter}}
\end{figure*}

In \autoref{sec:polarization}, we explore polarization spectral indices $\alpha_p$ and $\alpha_v$, how rapidly the fractional linear and circular polarization change as a function of frequency.  We plotted broad trends as a function of eDF in \autoref{fig:polarization_spectral_index}.  Now in \autoref{fig:polarization_spectral_index_scatter}, we provide full scatter plots of $\alpha_p$ and $\alpha_v$ for our M87* models between 214.1 and 350.0 GHz, using the plotting scheme introduced in \autoref{fig:alpha_tau_summary_M87_k5.0}.  Note the different extents of both axes in each panel.  We do not find any significant trends in these quantities as a function of spin, magnetic field state, or $R_\mathrm{high}$.  However, each eDF does cluster in a different region, as plotted in \autoref{fig:alpha_tau_summary_M87_k5.0}.

\section{Tests of the $\sigma_\mathrm{cut}$ Parameter}
\label{sec:sigma_cut}

As mentioned in \autoref{sec:grrt}, we zero all radiative transfer coefficients in regions with plasma $\sigma>\sigma_\mathrm{cut} = 1$.  This choice is frequently made when ray-tracing GRMHD studies due to numerical floors that artificially inject material in high $\sigma$ regions.  By construction, this results in the removal of otherwise emitting material near and within the jet.  While a complete survey of $\sigma_\mathrm{cut}$ is beyond the scope of this paper, we briefly gauge the impact of our choice of $\sigma_\mathrm{cut}=1$ by recomputing our M87* $\kappa=5$ models with $\sigma_\mathrm{cut}=10$.  We do not compute new values of $\mathcal{M}$ and merely adopt the same values computed for $\sigma_\mathrm{cut}=1$.

\begin{figure*}
  \centering
  \includegraphics[width=\textwidth]{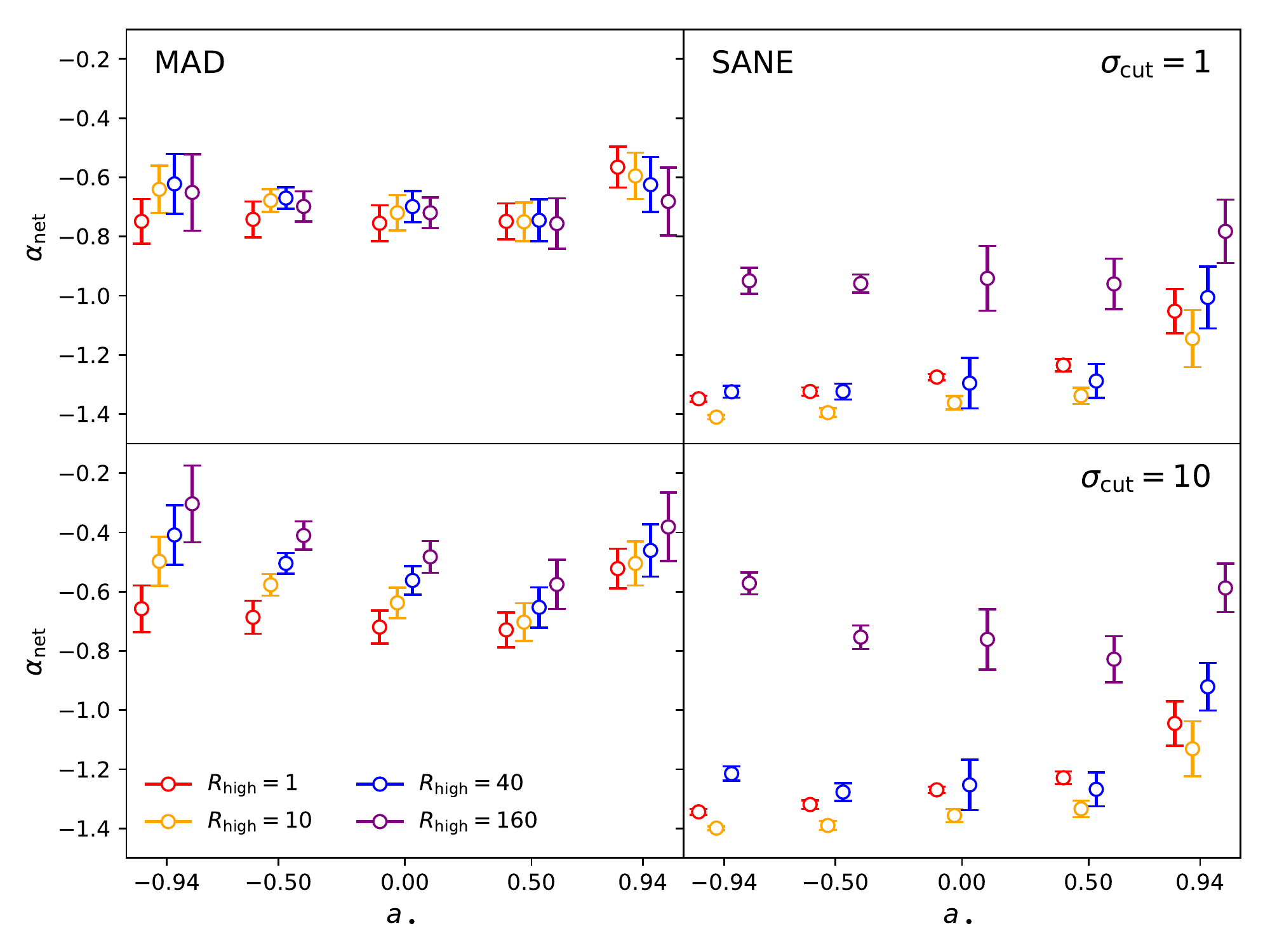}
  \caption{A comparison of models with $\sigma_\mathrm{cut}=1$ (top) and $\sigma_\mathrm{cut}=10$ (bottom) for our M87* $\kappa=5$ models in the style of \autoref{fig:unresolved_spectral_index_spin}.  When computing images with $\sigma_\mathrm{cut}=10$, we simply adopt the same values of $\mathcal{M}$ as used with $\sigma_\mathrm{cut}=1$.  Since more material is allowed to contribute to the emission, and this emission originates from hot magnetised regions, $\alpha_\mathrm{net}$ may increase.  The effect is larger for larger values of $R_\mathrm{high}$, since models with larger $R_\mathrm{high}$ also have larger $\mathcal{M}$, so the density in $\sigma>1$ regions is rescaled to be higher.  \label{fig:sigma_cut}}
\end{figure*}

A comparison of $\sigma_\mathrm{cut}=1$ and $\sigma_\mathrm{cut}=10$ models in the style of \autoref{fig:unresolved_spectral_index_spin} is shown in \autoref{fig:sigma_cut}.  While the general trends remain the same, the spectral index tends to become more positive with $\sigma_\mathrm{cut}=10$.  This is because more material is allowed to contribute to the emission, increasing the optical depth, and this additional emission also originates from hot magnetised regions.  Larger differences are seen for models with larger values of $R_\mathrm{high}$, which have larger $\mathcal{M}$.  Models where $\kappa$ varies based on local plasma properties \citep[e.g.,][]{Davelaar+2018,Fromm+2022,Scepi+2022} are likely to be more sensitive to the choice of $\sigma_\mathrm{cut}$.

\end{document}